\documentclass[journal=jpcbfk,manuscript=article,layout=twocolumn]{achemso}

\usepackage{hyperref}
\usepackage{multirow}
\usepackage{amsmath}
\usepackage{amsfonts}
\usepackage{graphicx}% Include figure files
\usepackage{dcolumn}% Align table columns on decimal point
\usepackage{bm}% bold math
\usepackage[utf8]{inputenc} 
\usepackage[T1]{fontenc}
\usepackage{mathptmx}
\usepackage[table]{xcolor} % Used to create different colors for every author
\usepackage{appendix}
\usepackage{booktabs}
\usepackage[version=4]{mhchem} % has easy commands for chemical formulas: \ce{FeCl2} renders well
\DeclareUnicodeCharacter{2212}{-}
\DeclareUnicodeCharacter{0301}{-}
\DeclareUnicodeCharacter{03B5}{-}
\usepackage{caption, makecell}
\usepackage{threeparttable}

\mciteErrorOnUnknownfalse

\newcommand{\mytoprule}{\toprule[0.125em]}
\newcommand{\mybottomrule}{\bottomrule[0.125em]}
\newcommand{\mymidrule}{\midrule[0.125em]}

\SectionNumbersOn

\setkeys{acs}{maxauthors=40}

\newcommand{\changelocaltocdepth}[1]{%
  \addtocontents{toc}{\protect\setcounter{tocdepth}{#1}}%
  \setcounter{tocdepth}{#1}%
}

\usepackage{bibunits}
\defaultbibliographystyle{achemso}
\defaultbibliography{bibliografia}
\author{Amrita Goswami}
\affiliation[University of Iceland]
{Science Institute and Faculty of Physical Sciences, University of Iceland, VR-III, 107 Reykjav\'{\i}k, Iceland}

\author{Samuel Blazquez}
\affiliation[Universidad Complutense de Madrid]
{Departamento de Qu\'{\i}mica F\'{\i}sica,
Facultad de Ciencias Qu\'{\i}micas, Universidad Complutense de Madrid,
28040 Madrid, Spain.}

\author{Lucía Fern{\'a}ndez-Sedano Vázquez}
\affiliation[Universidad Complutense de Madrid]
{Departamento de Qu\'{\i}mica F\'{\i}sica,
Facultad de Ciencias Qu\'{\i}micas, Universidad Complutense de Madrid,
28040 Madrid, Spain.}

\author{Eva Gonz{\'a}lez Noya}
\affiliation[Instituto de Química Física Blas Cabrera]
{ Instituto de Química Física Blas Cabrera,
CSIC, C/Serrano 119,
28006 Madrid, Spain.%\\This line break forced with \textbackslash\textbackslash
}%

\author{Hannes J\'{o}nsson}
\affiliation[University of Iceland]
{Science Institute and Faculty of Physical Sciences, University of Iceland, VR-III, 107 Reykjav\'{\i}k, Iceland}
\author{Jacobo Troncoso}
\affiliation[Universidade de Vigo]
{Departamento de F\'isica Aplicada, Universidade de Vigo, Escola de Enxeñaría Aeronaútica e do Espazo, E 32004, Ourense, Spain}

\author{Carlos Vega}
\affiliation[Universidad Complutense de Madrid]
{Departamento de Qu\'{\i}mica F\'{\i}sica,
Facultad de Ciencias Qu\'{\i}micas, Universidad Complutense de Madrid,
28040 Madrid, Spain.}
\email{cvega@ucm.es}

\title[An \textsf{achemso} demo]
{
    Viscosity as a Smoking Gun for Complex Formation in Solution: \ce{Fe^{2+}} and \ce{Mg^{2+}} Chlorides as Examples
} 

\abbreviations{MD, complex, viscosity}

\begin{document}

%%%%%%%%%%%%%%%%%%%%%%%%%%%%%%%%%%%%%%%%%%%%%%%%%%%%%%%%%%%%%%%%%%%%%
%% The "tocentry" environment can be used to create an entry for the
%% graphical table of contents. It is given here as some journals
%% require that it is printed as part of the abstract page. It will
%% be automatically moved as appropriate.
%%%%%%%%%%%%%%%%%%%%%%%%%%%%%%%%%%%%%%%%%%%%%%%%%%%%%%%%%%%%%%%%%%%%%
% \setlength{\fboxrule}{0 pt}
\begin{tocentry}
% TODO: replace with TOC in the correct dimensions 
\includegraphics[width = \linewidth]{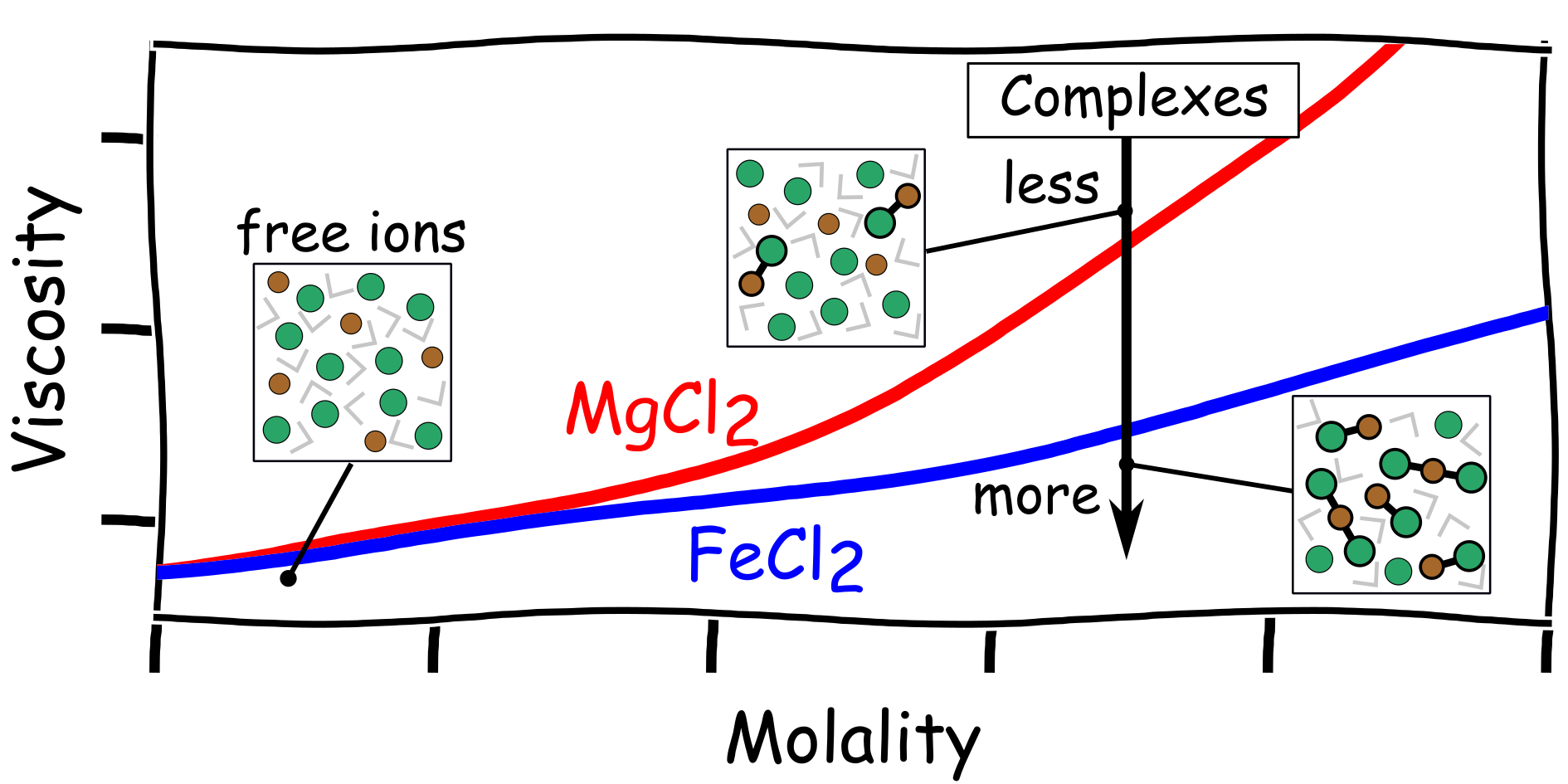}

\end{tocentry}

\setcounter{tocdepth}{-10}

%%%%%%%%%%%%%%%%%%%%%%%%%%%%%%%%%%%%%%%%%%%%%%%%%%%%%%%%%%%%%%%%%%%%%
% The abstract should not exceed 200 words
% Currently 185 words
\begin{abstract}
Electrolyte solutions at high concentration are indispensable and yet poorly understood. 
In particular, the extent of speciation -- the formation of complexes composed of multiple species -- in concentrated ionic solutions is very challenging to obtain theoretically and experimentally, but can have a strong effect on solution properties. 
The literature is rife with contradictory estimates of speciation from experiments. 
We find that speciation affects transport properties, and is therefore, a prerequisite to accurately model concentrated solutions. 
We turn this to our advantage by showing that the viscosity can be used to determine the extent of complexation in concentrated aqueous solutions. 
Results of simulations as well as experimental measurements are presented.
The atomistic Madrid-2019 force-field is extended to model \ce{FeCl2}.
Solutions of \ce{FeCl_2} and \ce{MgCl_2} are compared and the observed difference in viscosity explained by more complexation in the former, a conclusion supported by recently reported X-ray absorption and neutron scattering experiments. 
% Simulations using the Madrid-2019 force-field for the ferrous cation, \ce{Fe^{2+}}, in addition to \ce{Mg^2+} and \ce{Cl^-}, as well as complex distributions consistent with recently reported results of X-ray absorption and neutron scattering experiments, reproduce the experimentally observed trends. 
\end{abstract}

%%%%%%%%%%%%%%%%%%%%%%%%%%%%%%%%%%%%%%%%%%%%%%%%%%%%%%%%%%%%%%%%%%%%%
%% Introduction
%%%%%%%%%%%%%%%%%%%%%%%%%%%%%%%%%%%%%%%%%%%%%%%%%%%%%%%%%%%%%%%%%%%%%

\begin{bibunit}

\section{Introduction}
\label{sec:introduction}
Electrolytes in water are omnipresent in nature and play a vital role in practical and industrial applications.~\cite{Borodin2020,Khalid2023,Han2023,Luin2022a}  
For example, seawater is an electrolyte solution,~\cite{millero1974seawater} and many of the ions that are abundant in the sea are also present in living cells.~\cite{osterhout1935electrolytes} 
Given the ubiquity and relevance of electrolyte solutions, it is perhaps surprising that a rigorous and practical theory for electrolytes at high concentration is not available.
Many successful theories, such as the elegant Debye–Hückel theory,~\cite{debye1923theory} approach electrolyte solutions from the  infinitely dilute limit, and thus, quite intuitively, work only for dilute solutions.~\cite{Panagiotopoulos2020}
The behaviour of electrolyte solutions at high concentration is difficult to extrapolate from this infinitely dilute limit because of effects such as complex formation, ion pairing, etc., which cannot be easily shoehorned into a single mathematical framework.

The extent to which ions of opposite charge form long-lived ion pairs is, therefore, an important and divisive issue. 
The existence of complexes can significantly affect various properties of electrolyte solutions, particularly so at high concentration.\cite{hertz} 
However, experimental estimates differ widely on the extent of complexation, as well as on the equilibrium distribution of different types of complexes.
For example, values reported in the literature for the equilibrium constant, $\log K$, of \ce{FeCl^{+}} formation in aqueous \ce{FeCl_2} solution, range from $0.74$~\cite{wells_salam} to $-0.89$,~\cite{Boehm2015} with multiple values in between.\cite{olerup, iceland_team,Ruaya,Heinrich,Palmer,Zhao2001}  
Various measurements at $4$ m have, in particular, yielded contradictory results. 
The spectrophotometric experiments of \citet{Zhao2001} suggest that the monochloro complex \ce{[FeCl(H_2O)5]^+} is the predominant species in solution, with a small amount of the neutral trans dichloro complex \ce{[FeCl_2(H_2O)_4]}. 
However, X-ray absorption spectroscopy measurements of \citet{Luin2022} provide evidence for the conclusion that the neutral trans dichloro complex \ce{[FeCl_2(H_2O)_4]} is the dominant species.
In contrast, THz/FIR absorption spectra of \citet{Boehm2015} indicate the presence of only the monochloro complex. 

Chemical equilibrium models can predict speciation using activity models, for example Specific Ion Theory (SIT) \cite{sit_bronsted,sit_guggenheim,sit_ciavatta} and the Davies equation,~\cite{Davies1938} using ion association constants (typically obtained from experiments).~\cite{iceland_team, gustafsson2011visual}
However, the quality of the results depends strongly on (i) the activity model used, which may fail at high concentration, and (ii) the accuracy of the ion association constants.
Discrepancies can arise due to inherent differences in experimental measurement methods.
For instance, an equilibrium constant calculated using the anion exchange method,~\cite{Marcus1960} which is sensitive only to contact ion pairs, can be smaller than one estimated using potentiometry.~\cite{Tagirov2000} 

Atomic-scale simulations using
empirical potential functions can only approximately describe the interaction between molecules and ions. 
Despite this, they can still be used to learn some physics about electrolytes, thus extending our understanding of ionic solutions (at least qualitatively) beyond the Debye-Hückel limit. 
However, predicting both the dynamics of complex formation and the equilibrium distribution of complexes accurately can be challenging with simple point charge models.~\cite{DuboueDijon2017}

Density functional theory (DFT) has been successfully used to describe interactions between cations, water and anions in ionic solutions.\cite{zapalowski2001concentrated, dai2015ion, feng2017initial}
However, although quantum mechanical calculations can provide insight into the stability of specific complexes, they are infeasible to determine the equilibrium distribution of complexes due to their high computational costs and limited system size.

Therefore, the degree of complexation is difficult to ascertain from experiments, simulations and theory.
On the other hand, as discussed above, knowledge of complexation and complex distributions is crucial, since they can strongly affect solution properties. 
Here, we demonstrate that this problem can be turned on its head: viscosity can instead be used as a guide to infer the extent of complexation.

We accomplish this by performing both experiments, for \ce{FeCl_2}, and simulations, comparing \ce{FeCl_2} and \ce{MgCl_2}. 
In order to describe \ce{FeCl_2} in simulations, we extend the Madrid-2019 force-field\cite{zeron2019force},
which is a simple empirical force-field wherein monoatomic ions are represented by single Lennard-Jones (LJ) centers with a scaled charge.
We include complexes in our simulations by freezing distances between cation-chloride ion pairs. 
By explicitly including a fixed number of monochloro and dichloro complexes, and varying these constant proportions over a number of simulation trajectories, we show that the effect on the viscosity is surprisingly large and measurable.
We present clear evidence of higher association in \ce{FeCl_2}, compared to \ce{MgCl_2}.
This in contrast with the equilibrium values for association recommended by NIST,~\cite{NIST_SRD46} but which is in line with more recent experimental measurements.\cite{Luin2022,Callahan2010}

%%%%%%%%%%%%%%%%%%%%%%%%%%%%%%%%%%%%%%%%%%%%%%%%%%%%%%%%%%%%%%%%%%%%%
%% Experimental and Simulation details
%%%%%%%%%%%%%%%%%%%%%%%%%%%%%%%%%%%%%%%%%%%%%%%%%%%%%%%%%%%%%%%%%%%%%
\section{Methods}

\subsection{Experimental}
\label{subsec:expt_methods}
Ferrous chloride was purchased from Sigma-Aldrich, with a purity greater than $0.99$ in mass fraction. 
MilliQ water was used to prepare the solutions, made by weight in an AE-240 Mettler Toledo balance, with an uncertainty of $0.1$ mg, and in an atmosphere of \ce{N2} to avoid \ce{Fe^{2+}} oxidation.
Uncertainty in concentration is estimated to be $0.004$ mol/kg.
Densities were measured  using a DMA5000 vibrating tube densimeter from Anton Paar. 
Since the densities of the investigated solutions reach high values, calibration was performed using MilliQ water and carbon tetrachloroethylene, purchased from Sigma Aldrich with a purity greater than $0.999$ in mass fraction. 
The viscosity of the solutions is small enough to neglect any correction due to damping of the vibrating tube oscillations. 
Uncertainty in density is estimated to be $0.0003$ g/cm$^3$. 
More details about the procedure for density measurements can be found elsewhere.\cite{Sanmamed2009} 
Viscosities were measured using an AMVn falling ball viscometer, calibrated using MilliQ water, details of which are present in the literature.\cite{blazquez2025extending} 
Relative uncertainty in this magnitude is estimated to be $2 \%$. 
The experimental technique was validated by reproducing experimental densities and viscosities of \ce{MgCl2} from previous work.~\cite{Laliberte2004,laliberte2007model}
The density and viscosity  were determined for fourteen solutions at $298.15$ K, in the concentration interval ($0$-$4.1$) mol/kg.

\subsection{Computational}
\label{sec:simulation_methods}

\begin{table}[hbt!]
  \caption{Force-field parameters for the \ce{Fe^{2+}} and \ce{Cl^-} ions. The values for $q=0.85 \ e$ are those of the Madrid-2019 force field for \ce{Mg^2+}.
  % The parameters of \ce{Cl^-} for the model with $q=0.8$ \textit{e}  were taken from ~\citet{blazquez2023scaled}.
  The parameters which are different in the $q=0.80 \ e$ model, compared to the original model with $q=0.85 \ e$, are highlighted in bold text. 
  % In this work the same force field has been used for $FeCl_2$ and $MgCl_2$ (although of course the masses of Fe and Mg were different when running the simulations). 
  }
  \label{ff-parameters}
  \small
  \begin{tabular}{l c c}
      \mytoprule
      % \midrule[0.01em]
      & $q=0.85$ \textit{e} \\
      \midrule[0.05em] 
      & $\sigma_{ij}$ (nm) & $\epsilon_{ij}$ (kJ/mol) \\
      \midrule[0.05em] 
      \ce{Fe$^{2+}$-Fe$^{2+}$} & 0.116290 & 3.651900 \\[0.5em]
      \ce{Fe$^{2+}$-Cl$^{-}$} & 0.300000 & 3.000000 \\[0.5em]
      \ce{Fe^{2+}-$\mathrm{O_w}$} & 0.181000 & 12.00000\\[0.5em]
      \ce{Cl$^-$-Cl$^-$} & 0.469906 & 0.076923 \\[0.5em]
      \ce{Cl^--$\mathrm{O_w}$} & 0.423867 & 0.061983\\[0.5em]
      % \midrule[0.01em]
      \mymidrule
      % \midrule[0.01em]
      &  $q=0.80$ \textit{e} \\
      \midrule[0.05em] 
      & $\sigma_{ij}$ (nm) & $\epsilon_{ij}$ (kJ/mol) \\
      \midrule[0.05em] 
      \ce{Fe$^{2+}$-Fe$^{2+}$} & 0.116290 & 3.651900 \\[0.5em]
      \ce{Fe$^{2+}$-Cl$^-$} & 0.300000 & 3.000000 \\[0.5em]
      \ce{Fe$^{2+}$-$\mathrm{O_w}$} & \textbf{0.185000} & 12.00000 \\[0.5em]
      \ce{Cl$^-$-Cl$^-$} & 0.469906 & 0.076923 \\[0.5em]
      \ce{Cl$^-$-$\mathrm{O_w}$} & \textbf{0.418801} & 0.061983 \\[0.5em]
      % \midrule[0.01em]
      \mybottomrule
  \end{tabular}
\end{table}

Molecular Dynamics (MD) simulations were performed using both GROMACS (version 2025.2)~\cite{Abraham2015} and LAMMPS (29 Aug 2024, Update~1)~\cite{Thompson2022}; see Sec. 1.1.1 and Sec. 1.1.2 in the SI for details.
All results were obtained at ambient conditions (i.e., $298.15$ K and $1$ bar).
Atomic-scale simulations have been performed using the Madrid-2019 force-field,~\cite{zeron2019force} which combines the TIP4P/2005 model of water\cite{tip4p/2005} and scaled charges for the ions, such that monoatomic cations have a charge of $q=0.85 \ e$. The use of scaled charges (also denoted as the Electronic Continuum Correction) is becoming a  popular strategy in the literature.
\cite{leontyev09,leontyev10a,kirby2019charge,cruces2024building,jorge2019dielectric,le2020molecular,good_bad_hidden,Panagiotopoulos2020}

A version of the force-field with a charge of $0.80 \ e$ has also been used in this work.
% The canonical Madrid-2019 force-field has been used, wherein monoatomic cations have a charge of $q=0.85 \ e$, as well as a version of the force-field with a charge of $0.80 \ e$. 
Parameters for \ce{Mg^{2+}} (for $q=0.85 \ e$) and  \ce{Cl^{-}} (for both $q=0.85 \ e$ and $q=0.80 \ e$) have been reported previously~\cite{zeron2019force, blazquez2023scaled} whereas the parameters for  \ce{Mg^{2+}} (for $q=0.80 \ e$) were obtained in this work.
The proposed force-field parameters for \ce{Fe^2+} and \ce{Cl^-} are presented in Table~\ref{ff-parameters}.
% In fact one of the key ideas of this work is to use exactly the same force field for $MgCl_2$ and $FeCl_2$.
These parameters are the same as those of \ce{Mg^{2+}}.\cite{zeron2019force}  
Justifications for using the same force-field are provided in Sec.~\ref{subsubsec:madrid-19_fe_forcefield}.
Naturally, the correct masses should be used for \ce{Fe^{2+}} and \ce{Mg^2+}.

The molality scale, \emph{m}, has been used for concentrations. 
% (i.e. mol of salt per kg of water).
% Unless stated otherwise, the $0.85$ charge model has been used for all calculations.  

% However we found convenient to use also a force field with a different value of the scaled charge, namely $0.80 \ e$.  For this value of the charge the parameters of the interaction between chlorine and water were optimized in previous work. \cite{blazquez2023scaled}  In this work we optimized the interaction between $Mg^{2+}/Fe^{2+}$ and water for this value of the scaled charge. Again we used the same force field for $Fe^{2+}$ and $Mg^{2+}$ also for the model with the scaled charge $0.80 \ e$ (although of course we used different masses for Fe and Mg when doing the Molecular Dynamics runs). The parameters are presented in Table~\ref{ff-parameters}. 

%%%%%%%%%%%%%%%%%%%%%%%%%%%%%%%%%%%%%%%%%%%%%%%%%%%%%%%%%%%%%%%%%%%%%
%% Results
%%%%%%%%%%%%%%%%%%%%%%%%%%%%%%%%%%%%%%%%%%%%%%%%%%%%%%%%%%%%%%%%%%%%%
\section{Results}
\label{sec:results}

\subsection{ Measured density and viscosity of \ce{FeCl_2} }

\begin{table}[hbt!]
    \caption{Experimentally measured density and viscosity of \ce{FeCl_2} at various concentrations.  
     } 
    \label{jacobo}
    
    \begin{tabular}{l r r }
        \mytoprule
           Molality (mol/kg)    &  Density (g/$\mathrm{cm^3}$) &  $\eta$ (mPa$\cdot$s)  \\
        \midrule
        1.000 & 1.10188  & 1.286\\
        2.000 & 1.19758  & 1.846\\
        3.000 & 1.28578  & 2.663\\
        4.000 & 1.36809  & 4.170\\
        \mybottomrule
    \end{tabular}
\end{table}

Although the solubility limit of \ce{FeCl_2} is about $5.5$ m at room temperature and ambient pressure,~\cite{Chou1985} the literature lacks values for the density and viscosity for concentrations above $2$ m~\cite{Laliberte2004} and $0.4$ m,~\cite{laliberte2007model} respectively.
Data is reported here for up to $4$ m.
Interpolated values at round numbers for the density and viscosity are provided in Table~\ref{jacobo}. 
Good agreement is found with existing literature values for the more dilute solutions
[see also Fig. S2(a) in the SI].\cite{Laliberte2004,laliberte2007model}

\begin{figure}[H]
    \includegraphics[width=1.0\linewidth]{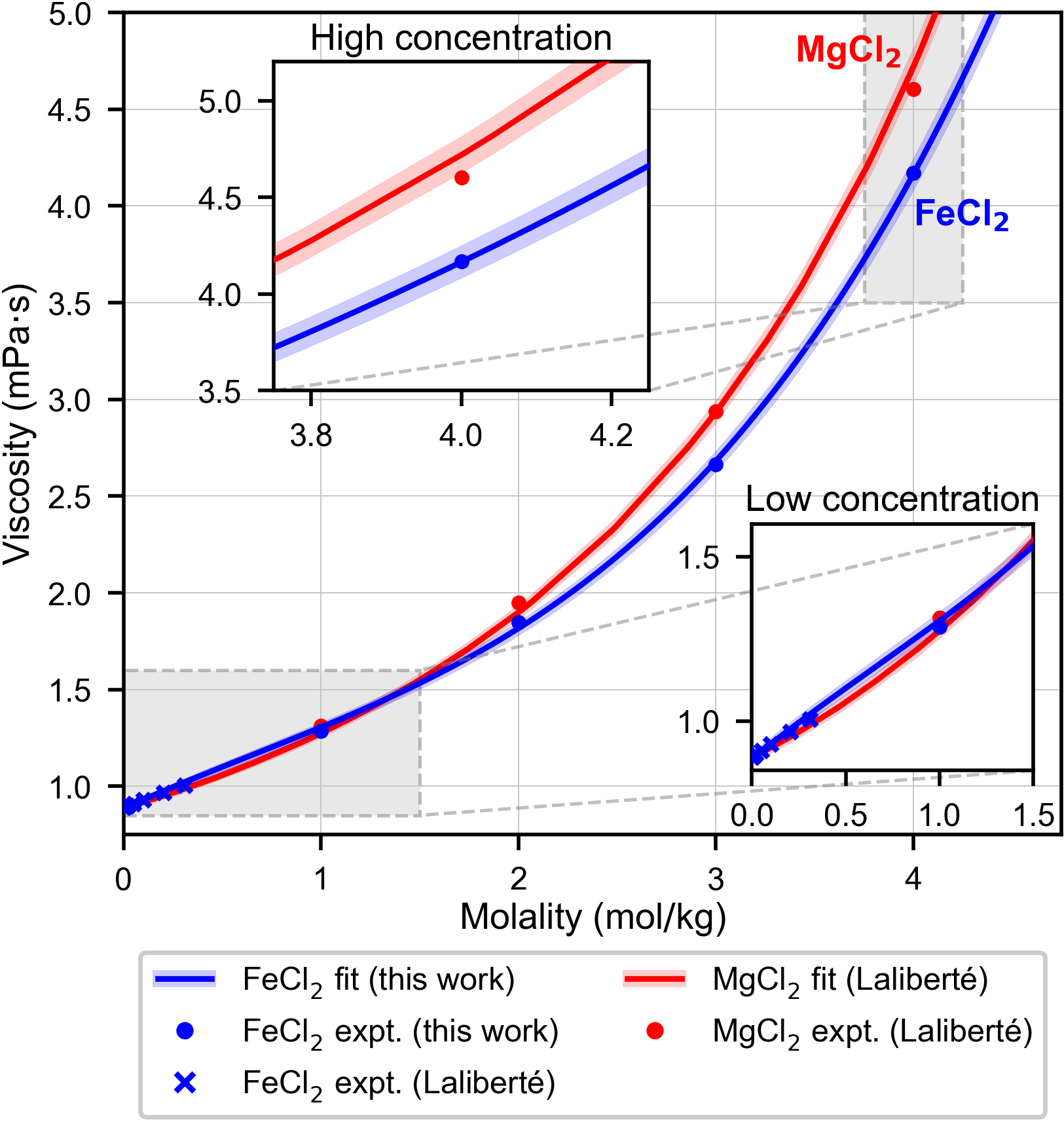}
    \caption{\label{fig:expt_viscosity}
    Viscosity of \ce{FeCl_2} and \ce{MgCl_2} from experiments, obtained from ~\citet{laliberte2007model} and this work. 
    The blue shaded region corresponds to a $2\,\%$ relative uncertainty in the \ce{FeCl2} viscosity measurements reported in this work.
    The red shaded region represents the $1.8\,\%$ mean relative deviation of the Laliberté model for \ce{MgCl2} from the experimental data~\cite{laliberte2007model}.
    % The shaded blue and red regions correspond to a relative error of $2 \%$, which is the uncertainty of viscosity measurements in this work for \ce{FeCl2}, and corresponds to a mean relative deviation from experimental viscosities for \ce{MgCl2}.
    Inset, bottom: Zoomed-in view of viscosities at low concentration, which are similar in value.
    Inset, top: Close-up view of the stark difference between the experimental viscosity of \ce{FeCl2} and \ce{MgCl2} at high concentration.
    } 
\end{figure}

 Fig.~\ref{fig:expt_viscosity} shows the experimental viscosities of \ce{FeCl2} and \ce{MgCl2} at both low and high concentration.
For concentrations less than $2$ m, the viscosities are similar, within errors (lower inset in Fig.~\ref{fig:expt_viscosity}).
The viscosity of electrolyte solutions is often described by the Jones-Dole equation:\cite{book-marcus} 
\begin{equation}
    \eta / \eta_w = 1 + A \sqrt{m} + B m + C m^2 
\label{jones_dole}
\end{equation}
where $\eta$ is the viscosity of the aqueous solution, $\eta_w$ is the viscosity of water, $m$ is the molality, the term $A$  [which is small~\cite{marcusb} and around $0.02$ (kg/mol)$^{1/2}$ for \ce{MgCl_2} and \ce{FeCl_2}] describes charge-charge contributions in the highly diluted region.
The $B$ coefficient (in units of kg/mol) is characteristic of individual ions, is additive, and can be interpreted in terms of ion-water interactions.~\cite{wang2004modeling}
The term $C$ [in units of (kg/mol)$^{2}$] is thought to depend on solute–solute association effects.~\cite{patil2014viscosity}
% and the term $C$ [in units of (kg/mol)$^{2}$] depends 
%on the interactions between hydrated ions.
% and the formation of complexes. 

Typically in experiments, the Jones-Dole coefficient $B$ is obtained by fitting the 
 viscosity at low concentrations (i.e., below $0.6$ m), where the quadratic term can be safely neglected. 
As can be expected from the similar viscosities of \ce{FeCl2} and \ce{MgCl2} at low concentration, the experimental values of $B$ for \ce{FeCl2} and \ce{MgCl2} are almost identical, equal to $0.405(10)$ kg/mol and $0.375(10)$ kg/mol, respectively (see also Sec.~\ref{subsec:no_complexes} for a comparison with simulations).~\cite{book-marcus} 
The value of $B$ for \ce{FeCl2} is slightly higher than that of \ce{MgCl2}, which could account for the corresponding marginally higher viscosity of the former compared to the latter at low concentration. 

However, the experimental viscosities of \ce{FeCl2} and \ce{MgCl2}, at high concentration, differ significantly, which is highlighted in the top inset of Fig.~\ref{fig:expt_viscosity}. 
% Eq.~\eqref{jones_dole} describes the viscosities well for lower concentrations, but is expected to fail at concentrations where ion pairs and complexes form.\cite{wang2004modeling}
Eq.~\eqref{jones_dole} describes the viscosities well up to a concentration of  $3$ m.
% , but is expected to fail at concentrations where ion pairs and complexes form.\cite{wang2004modeling}
We have estimated the values of $C$ as $0.12$ (kg/mol)$^{2}$ and $0.08$ (kg/mol)$^{2}$ for \ce{MgCl2} and \ce{FeCl2}, respectively, by fitting the experimental viscosities up to $3$ m. 
At higher concentrations, a cubic $D$ term [with units of (kg/mol)$^3$] would have to be added to Eq.~\eqref{jones_dole} to reproduce the experimental viscosities. 
% If the experimental viscosities upto $3$ m 
% In fact the value of $C$ when fitting experimental viscosities up to 3 m is 0.12 for $MgCl_2$ and 0.08 for $FeCl_2$ (for concentrations larger than 3.5 m a cubic term must be added to Eq.\ref{jones_dole} to reproduce the experimental viscosities).
% \amritaC{C coefficients}

The existence of monochloro and dichloro complexes in \ce{FeCl2} has been reported at concentrations around $4$ m.\cite{Zhao2001, Luin2022,Boehm2015}
On the other hand, \ce{MgCl2} being a strong 2:1 electrolyte, reportedly shows little or no complexation even at higher concentrations.\cite{DuboueDijon2017,bruni2012aqueous, Friesen2019}
In subsequent sections, we provide evidence from simulations that suggests that the difference of \ce{Fe^2+} and \ce{Mg^2+} in their propensity to complexation is tied to differences in their viscosity at high concentration, as already suggested by the large difference in the value of $C$ ($33\%$) between both systems.

% \subsection { A force field for $FeCl_2$ }
%  We mentioned previously that we will use the same force field in this work for $Fe^{2+}$ and $Mg^{2+}$.
%  It is interesting to analyze the consequences of that for volumetric (i.e densities) and transport properties ( viscosities ). 
 
%  \subsubsection{ $FeCl_2$ versus $MgCl_2$ : Densities }

\subsection{Simulations without complexes}
\subsubsection { A force-field for \ce{FeCl_2} } 
\label{subsubsec:madrid-19_fe_forcefield}

\begin{figure}[H]
    \includegraphics[width=1.0\linewidth]{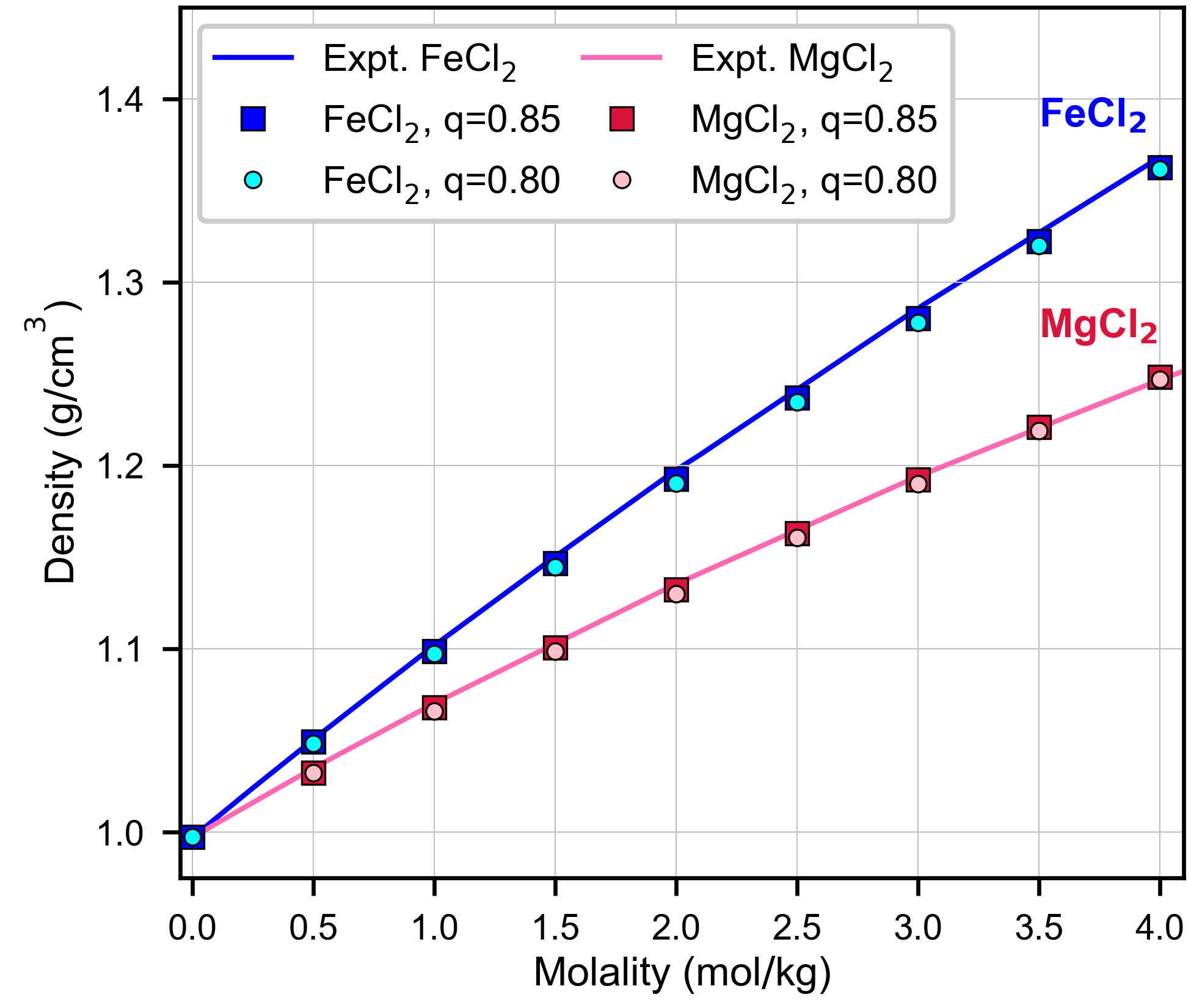}
    \caption{\label{fig:density_comparison}
    Density of \ce{FeCl2} and \ce{MgCl2} from fits to experiments (solid blue and magenta line, respectively) from this work and \citet{Laliberte2004, laliberte2007model}, respectively, as well as the density from atomic-scale simulations without complexes (squares and circles). 
    Agreement is within $0.5 \%$ for all concentrations.
    } 
\end{figure}

 Fig.~\ref{fig:density_comparison} presents a comparison of experimental densities with those obtained from simulations, using the same force-field for \ce{MgCl2} and \ce{FeCl2}, at various concentrations. 
 Agreement between simulations and experiments is within $0.5 \%$ for both the $0.85$ and $0.8$ charge models.
 Solution densities are often used as target properties in parameterization,\cite{zhang2017second, zeron2019force} and consequently, the good agreement with experiments at low and high concentration validates the approximation of using \ce{Mg^2+} force-field parameters for \ce{Fe^2+}.

 When the same force-field is used for \ce{FeCl2} and \ce{MgCl2}, an inherent assumption is that the number densities of both solutions are the same and that, consequently, the following relation holds true:

 \begin{equation}
    \rho_{\mathrm{FeCl_2}} = \rho_{\mathrm{MgCl_2}} \frac{ 1000 + m \times M_{\mathrm{FeCl_2}} }{1000+m \times M_{\mathrm{MgCl2}} },
\label{equal_volumes}
\end{equation}

where $\rho_{\mathrm{MgCl_2}}$ and $\rho_{\mathrm{FeCl_2}}$ are the mass densities in g/cm$^3$, $M_{\mathrm{MgCl_2}}$ and $M_{\mathrm{FeCl2}}$ are the molar masses in g/mol of \ce{MgCl_2} and \ce{FeCl_2}, respectively,
and $m$ is the molality of the solutions in mol/kg. 
 
 At low concentration, the assumption of equal number densities is a reasonable one, since the experimental values of the cation-oxygen distance and the hydration free energy of $\mathrm{Fe^{2+}}$ and $\mathrm{Mg^{2+}}$ are quite similar (collected in Table S1 in the SI).
In order to evaluate whether this assumption is also valid at higher concentrations, we estimated the mass density of \ce{FeCl2}, using Eq.~\eqref{equal_volumes}, from experimental \ce{MgCl2} densities, and compared the former with corresponding experimental densities, obtaining good agreement, as shown in Fig. S2(a) in the SI.
This simple test is also passed for \ce{Fe(SO_4)} and \ce{Mg(SO_4)}, as shown in Fig. S2(b) in the SI.
More generally, we believe that Eq.~\eqref{equal_volumes} can be used to easily test whether the force-field parameters of one cation can be reasonably used for another target cation (at least for density predictions) since such an evaluation would only require experimental densities at different concentrations.

\subsubsection{ Transport properties of solutions without complexes }
\label{subsec:no_complexes}
 % \subsubsection{ $FeCl_2$ versus $MgCl_2$ : Viscosities  }

 \begin{figure}[t]
    \includegraphics[width=1.0\linewidth]{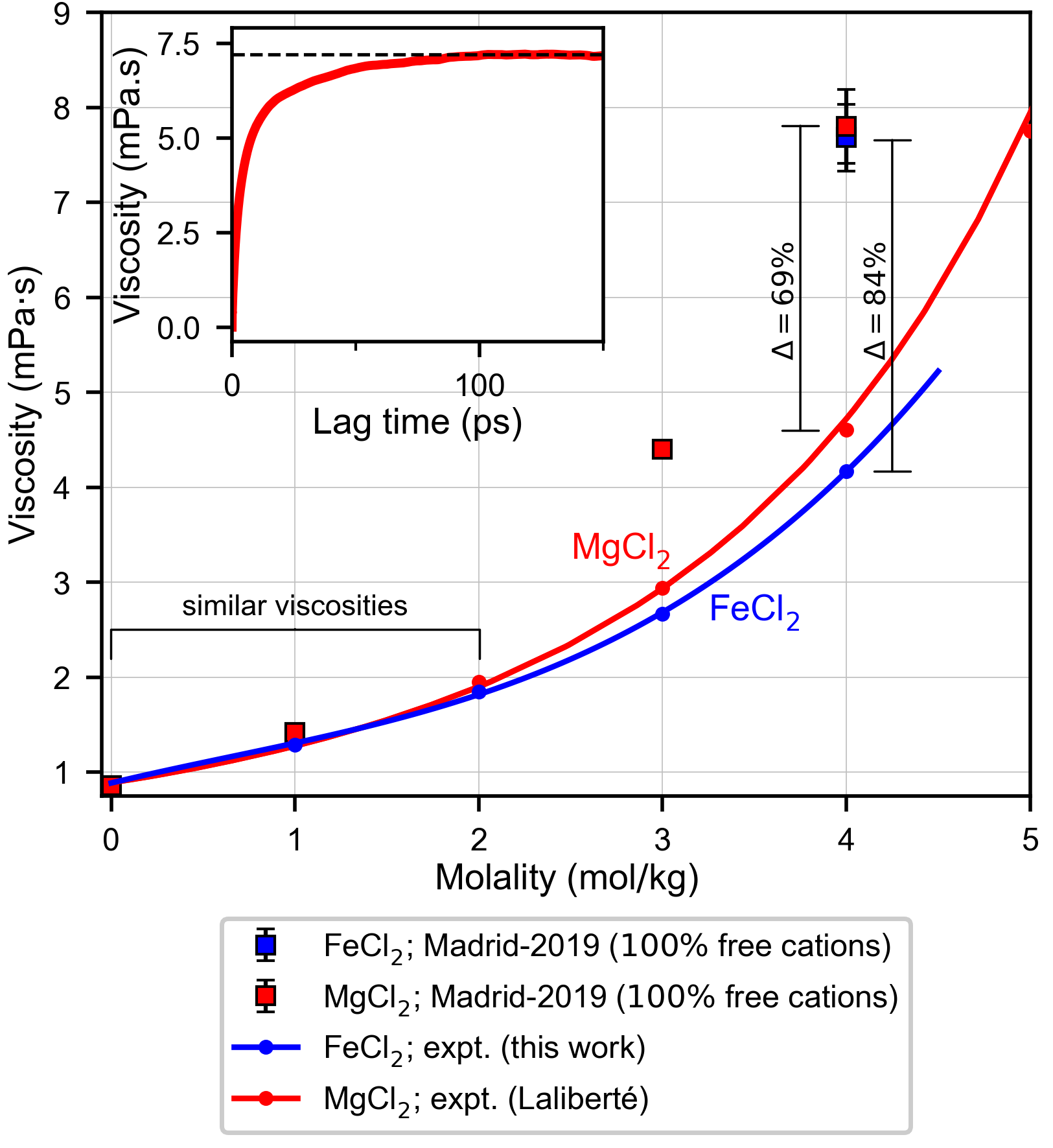}
    \caption{\label{fig:puzzle}
    Comparison of the viscosity of \ce{FeCl2} and \ce{MgCl2} obtained from simulations using the Madrid-2019 model (blue and red squares) with those from experiments (solid blue and red line, respectively).
    Inset: The viscosity obtained from the Green-Kubo formalism for a $4$ m \ce{MgCl2} solution, showing that the viscosity plateaus at about $100$ ps at the converged value.
    } 
\end{figure}

The experimental data obtained for \ce{FeCl2} enables us to examine trends in the viscosity at high concentration, where the effect of complexes is expected to be important.
However, complexes do not form spontaneously in our atomic-scale simulations of \ce{FeCl2} and \ce{MgCl2}.
% , when using the Madrid-2019 force-field.
This is because obtaining the equilibrium complex population would require much longer simulation time, on the order of microseconds, since this is the experimental residence time of water\cite{burgess,burgess_book2} in solvation shells of \ce{Mg^{2+}} or \ce{Fe^{2+}}.

 % This is in the SI already
% One of the consequences of having the same force-field parameters for \ce{FeCl_2} and \ce{MgCl2} is that, within the framework of classical statistical mechanics, they would have the same thermodynamic properties (i.e., the internal energies, radial distribution functions, enthalpies, vapor pressures, freezing point depression, boiling point etc.). 
% In contrast, transport properties could be different, in principle, owing to the fact that the mass of \ce{Fe^{2+}} is different from that of \ce{Mg^{2+}}.
% Nevertheless, the impact of the difference in the mass of the cations on global transport properties, such as the viscosity, is expected to be small in dilute electrolyte solutions. 

Fig.~\ref{fig:puzzle} shows the experimental values for the viscosity of \ce{FeCl_2} and \ce{MgCl_2}, along with results obtained from simulations (red and blue squares).
At concentrations below $2$ m, the viscosity of \ce{MgCl2} and \ce{FeCl2} obtained from simulations agree with experiments well, and this is also the regime where the experimental viscosities are quite similar (lower inset in Fig.~\ref{fig:expt_viscosity}).

\begin{table}[hbt!]
  \caption{
     Jones-Dole $B$ coefficients from simulations ($B_{\mathrm{sim}}$), computed using the $0.80$ Madrid-2019 model, and those from experiments ($B_{\mathrm{expt}}$) for \ce{FeCl2} and \ce{MgCl2}.\tnote{*}
    }
    \label{tab:jones_dole_sim}
  \begin{threeparttable}
    \small
    \begin{tabular}{l c c}
      \mytoprule
     System & $B_{\mathrm{sim}}$ (kg/mol) & $B_{\mathrm{expt}}$ (kg/mol) \\
      \midrule[0.01em]
       \ce{FeCl2} & $0.46(3)$ & $0.405(10)^{a}$ \\
       \ce{MgCl2} & $0.42(3)$ & $0.375(10)^{a}$ \\
      \mybottomrule
    \end{tabular}
    \begin{tablenotes}\footnotesize
      \item[*] a) From \citet{book-marcus}
    \end{tablenotes}
  \end{threeparttable}
\end{table}

Furthermore, at a concentration of $0.6$ m, it can be assumed that the quadratic term in Eq.~\eqref{jones_dole} is negligible (elaborated in Sec. S12 in the SI).
% This is confusing and also present in more detail in the SI. No need to have it here.
% (in fact, by using the experimental viscosity at $0.6$ m and neglecting the quadratic term, one obtains an estimate of $B$ which is only $0.01$ kg/mol higher than the reported value\cite{book-marcus}).
We estimated $B$ coefficients for the $0.80$ Madrid-2019 model by determining the viscosity at this concentration (see also Sec.~S12 in the SI for details of these calculations).
Table~\ref{tab:jones_dole_sim} presents these values, which are about $0.04$ kg/mol higher than experimental values.
Given that non-polarizable models with unscaled charges are known to overestimate
$B$ by approximately $0.16~\mathrm{kg\,mol^{-1}}$ for 1:1 electrolytes~\cite{yue2019dynamic},
with even larger deviations expected for 2:1 electrolytes, the level of agreement observed here is very good.
% Considering that non-polarizable models using non-scaled charges tend to overestimate $B$ for 1:1 electrolytes by $0.16$ kg/mol~\cite{yue2019dynamic} (for 2:1 electrolytes larger deviations are expected) this agreement is very good.

On the other hand, Fig.~\ref{fig:puzzle} shows that results from simulations deviate significantly from experiments at high concentrations around $4$ m.
The trends in Fig.~\ref{fig:puzzle} seem to suggest that the microstructure of \ce{FeCl_2} and \ce{MgCl_2} solutions is inherently different at high concentration, and that simulations which do not account for this fail to describe such solutions.
% Therefore, we acknowledge that, without explicitly parameterizing cation-chloride interactions to account for complex formation, it is not possible to describe the viscosity trends at high concentration, although such an approach has certain caveats as well (see Sec.\ref{subsec:limitations} for details).

% The trends in Fig.~\ref{fig:puzzle} seem to suggest a correlation between a difference in the propensity to form complexes and observed viscosity differences.
% We speculate that the microstructure of \ce{FeCl_2} and \ce{MgCl_2} solutions is inherently different at high concentration, compared to that at low concentration, and that simulations which do not account for this fail to describe such solutions. 
% Thus we recognize now that using the same force field for $FeCl_2$ and $MgCl_2$ exhibit limitations in describing the viscosity at high concentrations. However, we now turn to a new strategy. We will introduce complexes in the simulations, and will allow to have different amount of them for $FeCl_2$ and $MgCl_2$.

\subsection{Simulations with complexes}
%\subsection{ A simulation strategy for complexes }
%\label{subsec:complexes}

As touched upon previously in Sec.~\ref{subsec:no_complexes}, the formation of complexes is not observed even after $200$ ns runs with the Madrid-2019 force-field.
Additional interactions would need to be included to capture complex formation.~\cite{DuboueDijon2017}
We conclude that both the long timescale for attaining the equilibrium complex distribution in concentrated solutions, and the
challenge of describing the interaction between the ions at such close range,
make simulations describing the dynamics of complex formation impractical at this time (see Sec.~\ref{subsec:limitations} for an expanded discussion).

Here, we make a distinction between the somewhat synonymous terms -- contact ion pairs (CIPs) and complexes -- used in the literature. 
We use the term CIP when the residence time of the anion in contact with the cation is relatively low (from a few to dozens of picoseconds), such as in the case of NaCl.~\cite{Elliott2024}
We assume that the lifetime of such CIPs is less than the typical decay time of the stress autocorrelation function for the solution (which is around $100$-$200$ ps for the systems considered in this work as shown in the inset of Fig.~\ref{fig:puzzle}).
In this work, we use the term complex when the residence time of the anion in contact with the cation is  significantly larger than the decay time of the stress autocorrelation function, so that  we can assume that their distribution does not change significantly during the course of a simulation trajectory used to calculate physical quantities of interest (here, the viscosity and diffusion coefficient).

% \subsubsection{ Worked complex examples: monomers and dimers }
\subsubsection{A simulation strategy for including complexes }
In our simulations, we model complexes by freezing the cation-chloride distance to $2.33$ \AA ~(see Sec.~4 of the SI for justifications)
Concomitantly, water molecules are allowed to move, and are free to translate, rotate, and even leave the solvation shell. 
Freezing the cation-anion distances is a reasonable approximation, since 
cation-chloride vibrations within the complex 
are not expected
to have a significant effect on the viscosity or diffusion coefficient.

\begin{figure}[H]
  \includegraphics[width=1.0\linewidth]{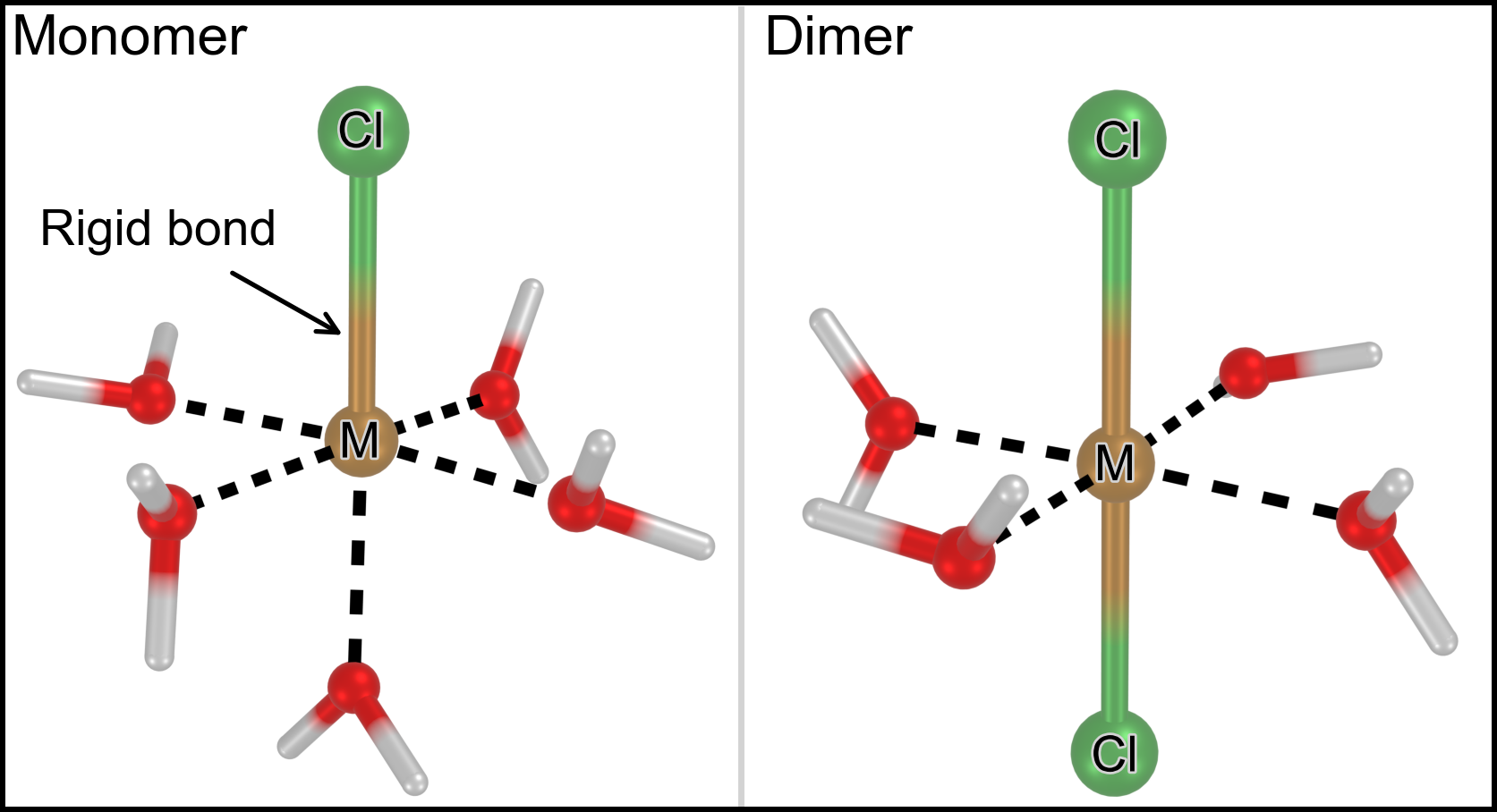}
  \caption{
  Illustration of the complexes considered in the atomic-scale simulations. 
  The bond between the metal cation and \ce{Cl-} anion is rigid, but the water molecules are free to move. 
  The black dashed lines indicate the octahedral shape of the solvation shell. 
  \emph{Left:} The \emph{monomer} complex, wherein the cation is part of a 
monochloro complex (where \ce{M} corresponds to an \ce{Fe^2+} or \ce{Mg^2+} cation).   
  \emph{Right:} The \emph{dimer} complex, consisting of a linear 
  dichloro unit \ce{Cl-M-Cl}. 
  }
  \label{free_bound_ions_illustration}
\end{figure}

Complexes can also exhibit a plethora of structures, which is why speciation generally refers to a distribution of these types.
We use the following terminology to describe two of the most dominant types of complexes formed by \ce{Fe^2+}~\cite{Zhao2001,Boehm2015,Luin2022} and \ce{Mg^2+},~\cite{Friesen2019,pye1998ab} illustrated in Fig.~\ref{free_bound_ions_illustration}.
A \emph{monomer} (Fig.~\ref{free_bound_ions_illustration} left) refers to a complex formed by one cation, coordinated with one \ce{Cl^{-}} anion and some additional molecules of water in the first solvation shell, with the cation-chloride distance frozen.
A \emph{dimer} (Fig.~\ref{free_bound_ions_illustration} right) refers to a complex wherein there are two \ce{Cl^{-}} anions in the trans position of the octahedron, along with additional coordinating water molecules.

In our new simulation strategy, instead of attempting to model the dynamics of complex formation, we \emph{introduce a fixed number of complexes} (monomers and/or dimers) at the beginning of simulations as an input 
and constrain the cation-chloride distance in the complex to study the effect that complex distributions can have on the properties of the solutions.

This strategy was validated by analyzing the structures of the solvation shell around free cations, monomers and dimers. 
The octahedral solvation structure was retained by monomers and dimers, showing that constraining the cation-chloride distance does not distort the solvation shell around the cation (Fig. S3 in the SI). 

In addition, the effect of complexes on the density is small, at least when considering only octahedral complexes (see Table S3 in SI; the density increases by only $0.5\%$ at $4$ m when all cations participate in monomers and there are no free cations).
Therefore, the introduction of monomers and dimers does not cause the excellent density predictions of the force-field to deteriorate. 

\subsubsection{The effect of complexes on transport properties}
\label{subsec:complex_effects}
 
We denote $\alpha$ as the percentage of free cations, $\alpha'$ as the percentage of monomers and $\alpha''$ as the percentage of dimers, such that:
\begin{equation}
    \label{normalization_1}
    \alpha + \alpha ' + \alpha'' = 100. 
\end{equation}

First, we analyze the impact of monomers (illustrated in Fig.~\ref{free_bound_ions_illustration}) on the viscosity.
When considering only free ions and monomers, $\alpha ''$ is zero.
We vary the fraction of free cations, $\alpha$, from $100 \%$ (where all ions are free) to $0 \%$ (where all cations form monomers). 

\label{subsec:transport_complex_dependence}
\begin{table}[hbt!]
    \caption{Viscosity (in mPa$\cdot$s)  of \ce{FeCl_2} and \ce{MgCl_2} solutions at $4$ m obtained from simulations using the Madrid-2019 force field. 
    % $\alpha$ denotes the percentage of free cations in the system (only monomers are considered so that $\alpha+\alpha'=100$). 
    The Yeh-Hummmer correction~\cite{Yeh2004} was used to account for finite-size effects.
    % The Yeh-Hummer correction~\cite{Yeh2004} has been applied to the diffusion coefficients to account for finite-size effects. 
    } 
    \label{viscosity_with_without_monomers}
    \small
    \begin{tabular}{lccc}
        \mytoprule
        System & $\alpha$ & $D_{\ce{H_2O}}$ $\times 10^{9}$  ($\mathrm{m^2/s}$) & $\eta$ (mPa$\cdot$s)  \\
        \midrule
        \ce{MgCl_2} & $100$ & $0.366$ & $7.73(0.39)$ \\
        \ce{MgCl_2} & $15$  & $0.499$ & $5.71(0.33)$ \\
        \ce{MgCl_2} & $2.5$ & $0.522$ & $5.56(0.42)$ \\
        \ce{FeCl_2} & $100$ & $0.369$ & $7.68(0.41)$ \\
        \ce{FeCl_2} & $15$  & $0.496$ & $5.87(0.30)$ \\
        \ce{FeCl_2} & $2.5$ & $0.522$ & $5.54(0.32)$ \\
        \mybottomrule                             
    \end{tabular}
\end{table}

\begin{figure}[t]
    \centering
    \includegraphics[width=1.0\linewidth]{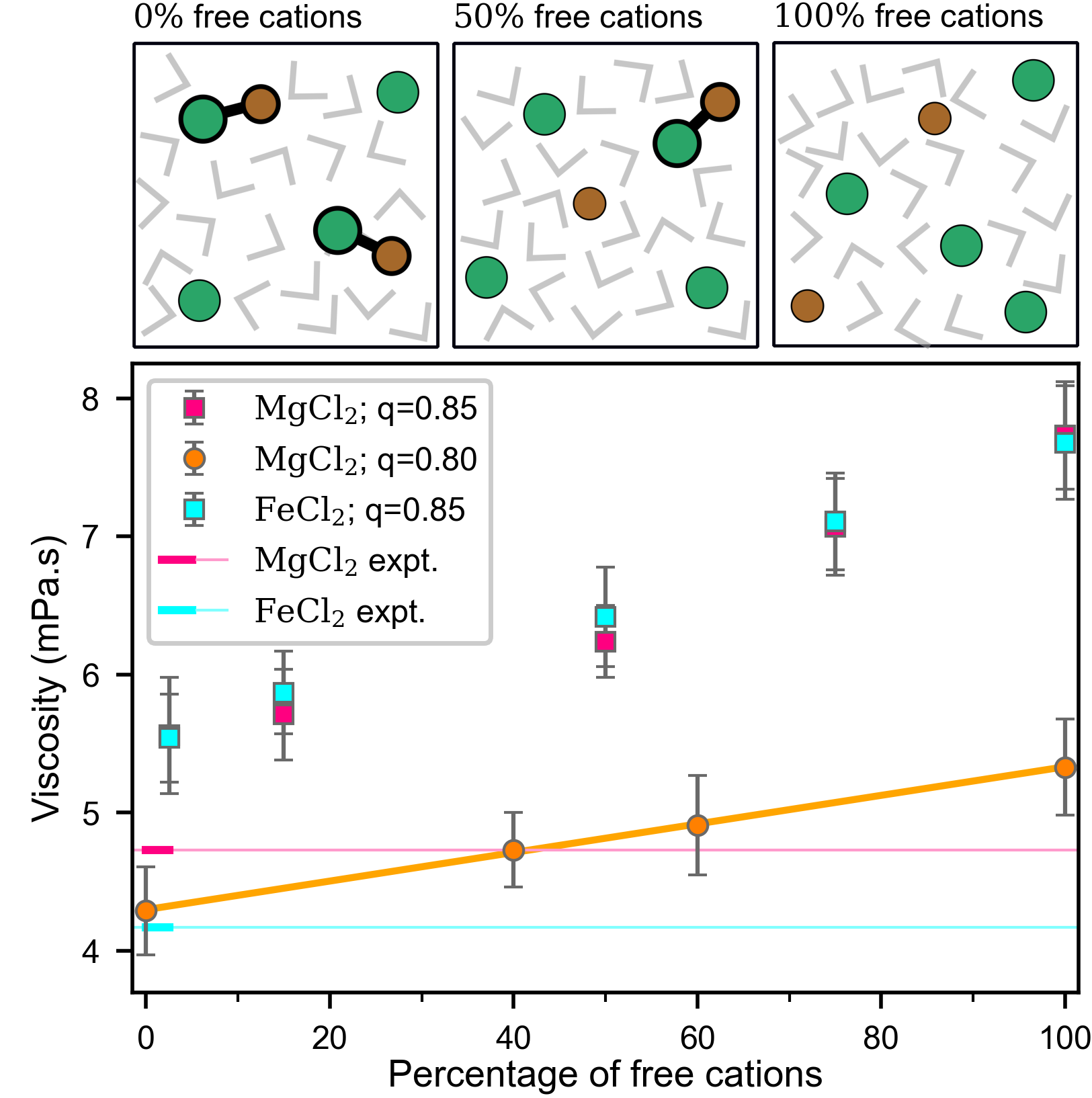}
    \caption{The experimentally measured viscosity of a $4$ m solution of \ce{MgCl_2}~\cite{laliberte2007model} and \ce{FeCl_2} depicted as magenta and cyan horizontal lines, respectively,
    as well as the 
    calculated viscosity (squares) of 
    solutions containing a varying fraction of monomer complexes and free ions, shown as a function of the percentage of free cations, $\alpha$.
    The orange circles depict calculated viscosity of \ce{MgCl_2} using the Madrid-2019 force-field with a charge of $0.80 \ e$. 
     Insets on top illustrate solutions with 
    no free cations, 50\% free cations and only free cations. }
    \label{fig:visc_change_with_monomers_4m}
\end{figure}

The viscosity and diffusion coefficient of water obtained in simulations for $4$ m solutions of \ce{FeCl_2} and \ce{MgCl_2}
are presented in Table~\ref{viscosity_with_without_monomers}.
Fig.~\ref{fig:visc_change_with_monomers_4m} shows the viscosities of \ce{MgCl_2} 
%(magenta squares) 
and \ce{FeCl_2} 
%(cyan squares) 
at $4$ m (at $298.15$ K and $1$ bar), for different complex distributions (i.e., different proportions of free cations and monomers). 
The experimental viscosities of \ce{MgCl_2} and \ce{FeCl_2} at this concentration are also depicted.
%as light magenta and cyan lines.  
We observe that the viscosities of \ce{FeCl_2} and \ce{MgCl_2} from simulations are roughly the same (within our uncertainty, which is $0.25$-$0.45$ mPa$\cdot$s). 
Notably, the presence of complexes (monomers, in this case) significantly reduces the value of the viscosity -- the viscosity of \ce{MgCl_2} goes from $\approx 7.73$ mPa$\cdot$s when $\alpha=100$  to $\approx 5.5$ mPa$\cdot$s when $\alpha=0$.
Note that this last value of the viscosity, $5.5$ mPa$\cdot$s, is closer to the experimental result ($4.73$ mPa$\cdot$s).~\cite{laliberte2007model}
%As shown in Table \ref{viscosity_with_without_monomers}, t
% The diffusion coefficient of water increases significantly and the viscosity decreases as the number of free ions decreases.
%As the fraction of complexes increases, the viscosity decreases.
%The explanation of this trend is intuitive: ions tend to slow down the dynamics of the solvated water molecules associated with them.~\cite{Shi2023}
% water molecules 
%by solvating them.
% because of the solvation.
%Additionally, the net charge of a monomer is half that of a free cation, thereby exerting a weaker electric field on surrounding water molecules.
%Thus, the presence of complexes (here monomers) reduces the net amount of encumbered water molecules in contact with the ions, compared to solutions with free ions. 

The explanation of the decrease of viscosity with increase in complexes is intuitive: water molecules in the vicinity of ions diffuse more slowly than water molecules in the bulk.~\cite{Shi2023}
% water molecules in the vicinity of ions diffuse more slowly than free water molecules.~\cite{Shi2023}
% ions tend to slow down the dynamics of the solvated water molecules associated with them.~\cite{Shi2023}
Additionally, the net charge of a monomer is half of that of a free cation, thereby exerting a weaker electric field on the surrounding water molecules.
Thus, the presence of complexes (monomers, in this case) reduces the net amount of encumbered water molecules in contact with the ions, compared to solutions with free ions. 

Consequently, the diffusion of water increases with a rise in the number of complexes, as shown in Table \ref{viscosity_with_without_monomers}. 
The observation that the viscosity and the diffusion coefficient of water are anti-correlated in electrolyte solutions was discussed and shown by \citet{mccall1965} more than sixty years ago. 
This suggests that experimental measurements of $D_{\ce{H_2O}}$ could also serve as an indirect measure of complex formation in these systems. 

% moved to section above where we discuss validation of complex addition strategy
% Notably, although the viscosity is quite sensitive to the presence of complexes (as shown in Table \ref{viscosity_with_without_monomers}) the effect of complexes on the density is negligible (see SI for further evidence\amritaC{which section and table}), at least when considering only octahedral complexes.
% Thus the introduction of monomers and dimers does not deteriorate the excellent density predictions of the force-field.  

However, even after introducing monomers, quantitative agreement with experiments is not achieved. 
Fig.~\ref{fig:visc_change_with_monomers_4m} shows that even if we assume $100\%$ population of monomers and no free cations, neither the viscosity of \ce{MgCl_2} nor that of \ce{FeCl_2} is reproduced. 
Of course, one should not disregard the presence of other complexes in the system, for instance those formed by the \ce{FeCl_2} species (or the dimer species, in our terminology, see Fig.~\ref{free_bound_ions_illustration}).  

Since dimers presumably hinder even less water molecules than monomers, we surmise that the presence of dimers should reduce the viscosity further. 
At $4$ m  with $\alpha''=100$ (all the cations form dimers), the viscosity of the  system (either \ce{MgCl_2} or \ce{FeCl2}) is calculated to be $4.8(0.2)$ mPa$\cdot$s, which is lower than that of the system wherein all complexes form monomers [$\approx5.5$  mPa$\cdot$s]. 
% Thus, dimers reduce the viscosity even more than monomers.
However, this reduction is still not low enough to reach the experimental viscosity of \ce{FeCl_2}, which is $4.17$ mPa$\cdot$s (as shown in Table ~\ref{jacobo}).

% Although scaled charge models improve the description of transport properties with respect to models using formal charges, the Madrid-2019 (with a scaled charge of $0.85 \ e$) still overestimates the viscosity even for simple salts such as NaCl, KCl, etc (see also Sec. 9 of the SI).
% It has been demonstrated recently that reducing the value of the scaled charge to a lower value improves the description of transport properties.\cite{blazquez2020scaled, pavel_2025} 
% Therefore, we also analyze the results of simulations with the $0.8$ charge model. 
% Notice that not only the charge is modified but also the LJ parameters (see Table \ref{ff-parameters}).
% With this new value of the scaled charge 
We also analyze the results of simulations with the $0.8$ Madrid-2019 model, which tends to improve the description of transport properties, compared to the $0.85$ model.\cite{blazquez2020scaled, pavel_2025} 
Using this model, it is possible to reproduce the experimental viscosity of both \ce{MgCl2} and \ce{FeCl2}, by assuming $40\%$ of free cations for \ce{MgCl_2}  and $0\%$ for \ce{FeCl_2} (i.e., all cations participate in monomers), as shown in Fig.~\ref{fig:visc_change_with_monomers_4m} (Table S7 in the SI presents simulation results with $100\%$ dimers). 
Our results indicate a $35$–$40$\% greater degree of complex formation in \ce{FeCl2} compared to \ce{MgCl2}.
Therefore, we surmise that the association for \ce{FeCl_2} is significantly
more important than for \ce{MgCl_2}, a conclusion supported by recent experiments.\cite{Luin2022,Callahan2010}

%%%%%%%%%%%%%%%%%%%%%%%%%%%%%%%%%%%%%%%%%%%%%%%%%%%%%%%%%%%%%%%%%%%%%
%% Discussion
%%%%%%%%%%%%%%%%%%%%%%%%%%%%%%%%%%%%%%%%%%%%%%%%%%%%%%%%%%%%%%%%%%%%%
\section{Discussion}
\label{subsec:discussion}

\subsection{Complex populations:  the thermodynamic formalism }
\label{subsec:inconsistencies}

Although experiments of \ce{FeCl2} solutions unanimously report the existence of complexes at high concentration, there is little consensus on the nature and type of dominant complexes formed.~\cite{Zhao2001,Boehm2015,Luin2022} 
We speculate that determining their precise equilibrium distribution can be quite difficult from experiments, and, to the best of our knowledge, quantitative complex distributions have not been reported so far. 

Another conceptually attractive route to estimating complex populations could be to obtain equilibrium constants.
We define the equilibrium constant $K_1$ for the reaction $\ce{M^{2+}} +  \ce{Cl^{-}} \rightleftharpoons \ce{MCl^{+}}$ as: 
\begin{equation}
    K_1 = \frac{ a_{\ce{MCl^+}} }{ a_{\ce{M^{2+}}} \; a_{\ce{Cl^-}} } = \frac{ \ce{[MCl^{+}]} }{ \ce{[M^{2+}]} \ce{[Cl^{-}]} } \frac{\gamma_{\ce{MCl^+}}} {\gamma_{\ce{M^{2+}}} \; \gamma_{\ce{Cl^-}}},
    \label{def_K1}
\end{equation}
where \ce{M} refers to the metals considered here (\ce{Fe} or \ce{Mg}), $a_{\mathrm{i}}$ is the activity of component $i$, and $\gamma_{\mathrm{i}}$ is its activity coefficient, and [$i$] is the concentration of component $i$.

However, there is a wide disparity in the value of the estimated equilibrium constant 
$K_1$.
Table \ref{chaos_in_values_of_K} presents $\log K_1$ from the literature, which varies from $0.74$ to $-0.89$ for \ce{FeCl2}, corresponding to a difference of almost two orders of magnitude. 
The value of $\log K_1$ of \ce{FeCl2} recommended by NIST is $-0.2$, which lies squarely in the middle of the range shown in Table \ref{chaos_in_values_of_K}.
The same is true for \ce{MgCl2}, although it has been studied much less than \ce{FeCl2},~\cite{iceland_team, gustafsson2011visual} and the general consensus is that \ce{MgCl2} has low propensity to form complexes.\cite{DuboueDijon2017,bruni2012aqueous, Friesen2019}

% Another possible route to determine the population of complexes is to use thermodynamics. After all the formation of complexes depends on the equilibrium constants, and if they were know that would help considerably. 
% We define the equilibrium constant $K_1$ for the reaction $\ce{M^{2+}} +  \ce{Cl^{-}} \rightleftharpoons \ce{MCl^{+}}$ as: 
% \begin{equation}
%     K_1 = \frac{ a_{\ce{MCl^+}} }{ a_{\ce{M^{2+}}}  a_{\ce{Cl^-}} } = \frac{ \ce{[MCl^{+}]} }{ \ce{[M^{2+}]} \ce{[Cl^{-}]} } \frac{\gamma_{\ce{MCl^+}}} {\gamma_{\ce{M^{2+}}} \gamma_{\ce{Cl^-}}},
%     \label{def_K1}
% \end{equation}

  % The first problem is the value of the thermodynamic equilibrium constant 
  % $K_1$. Values of $K_1$ reported in the literature are shown in 
  % Table \ref{chaos_in_values_of_K}. As you can see values, in the case 
  % of $Fe^{2+}$ varies from 0.74 to -0.89. Notice that these are log terms, so that
  % there are almost two orders of magnitude of difference between estimates of different authors. Which value is the correct one ?
  % NIST recommends -0.2 (a value in the middle range of those reported in Table \ref{chaos_in_values_of_K}). Att this point we can not state with certainty which value of the equilibrium constant is correct. The same is true for $MgCl_2$ which has been study much less often than $FeCl_2$.~\cite{iceland_team, gustafsson2011visual}

\begin{table}[hbt!]
    \caption {
        Equilibrium  constants $\mathrm{log}(K_1)$ for monomer formation of the reaction 
        $\ce{M^{2+}} + \ce{Cl^{-}} \ce{<-->  MCl^{+}}$, where \ce{M} refers to the metal. 
    }
    \label{chaos_in_values_of_K}
    \small
    \begin{tabular}{lr}
        \mytoprule
        \textbf{\ce{FeCl_2}} \\
        \midrule
        H. Olerup   (1944)~\cite{olerup}              &                      $0.36$   \\
        Wells and Salam (1967)~\cite{wells_salam}           &                     $0.74$  \\
        Arnorsson et al.(1982)~\cite{iceland_team}             &                 $-0.40$   \\
        J.R.Ruaya      (1988)~\cite{Ruaya}                             &    $-0.50$  \\
        Heinrich and Seward (1990)~\cite{Heinrich}   &     $-0.16$  \\
        Palmer and Hyde  (1993)~\cite{Palmer}        &     $-0.125$ \\
        Zhao and Pan   (2001)~\cite{Zhao2001}                             &     $-0.366$ \\
        Böhm et al. (2015)~\cite{Boehm2015}                           &    $-0.89$  \\
        NIST               &                                   $-0.2$  \\
        \mymidrule
        \textbf{\ce{MgCl_2}} \\
        \midrule
        J.R.Ruaya (1988)~\cite{Ruaya}             &            $-0.13$    \\
        NIST                  &             $0.6$ \\
        \mybottomrule
    \end{tabular}
\end{table}

However, quantifying complex populations involves another contentious issue: activities (for instance, $a_{\ce{MCl^+}}$, $a_{\ce{M^{2+}}}$, $a_{\ce{Cl^-}}$), and not concentrations, need to be estimated for real solutions. 
Individual activity coefficients cannot be directly calculated from experiments.
Activity models, such as as Specific Ion Theory (SIT) \cite{sit_bronsted,sit_guggenheim,sit_ciavatta} or the Davies equation~\cite{Davies1938} are often used to estimate complex populations, given equilibrium constant values.
Chemical equilibrium models, such as Visual MINTEQ (version 4.0),~\cite{gustafsson2011visual} can estimate percentage distributions of free cations, using such activity models.
However, even when starting from the same equilibrium constant (for instance, those recommended by NIST), different activity models yield widely diverging complex populations, as shown in Table~\ref{free_cations_vmintec_4_0}. 

\begin{table}[hbt!]
  \caption{ Percentage of free cations, $\alpha$, in $4$ m aqueous solutions at room temperature and pressure, obtained from Visual MINTEQ~\cite{gustafsson2011visual} using different activity models and association constants from NIST.  
  % This program uses the association constants recommended by NIST.
  The column labeled $\gamma=1$ refers to results obtained from Eq.~\eqref{def_K1} using the approximation that all activity coefficients are unity.
  More association is predicted for \ce{MgCl_2} than for \ce{FeCl_2}.} 
  \label{free_cations_vmintec_4_0}
  \small
  \begin{tabular}{l c c c c}
      \mytoprule
      \multirow{2}{*}{System} & \multicolumn{3}{c}{Activity model} \\
      \cmidrule(lr){2-4}
      & SIT & Davies &  $\gamma=1$\\
      \midrule
      \ce{MgCl_2} & $8\%$ & $0.5\%$ & $6\%$\\ 
      \ce{FeCl_2} & $29\%$ & $3\%$ & $24\%$ \\
      \mybottomrule
  \end{tabular}
\end{table}

As can be expected, calculations using such chemical equilibrium models are sensitive to the quality of the activity model and the ion association constants. 
We surmise that, at this point, they are not reliable enough to unambiguously determine the concentration of complexes.
  
\subsection{Toy model exploiting the relationship between complex populations and viscosity}

In Sec.~\ref{subsec:complex_effects}, we presented a model that could reproduce the experimental viscosities of \ce{FeCl_2} and \ce{MgCl_2} simultaneously, contingent on a particular (variable) complex population used to fit to the viscosity (solid orange line in Fig.~\ref{fig:visc_change_with_monomers_4m}). 
% In fact the orange line of Fig.\ref{fig:visc_change_with_monomers_4m} would reproduce the experimental viscosities if one one assume , that there are no dimmers, and that the number of free ions is 0 for $FeCl_2$ and of about $40 \%$ for $MgCl_2$. However there is some evidence that dimers are also found experimentally in $FeCl_2$.

Simulation results notwithstanding, we cannot claim to quantitatively determine the equilibrium concentration and the type of complexes that actually exist experimentally in \ce{FeCl_2} and \ce{MgCl_2}. 
% Even if only monomers and dimers are considered, there could be multiple possible complex populations that would satisfy the condition of viscosities that approach experimental values.
However, we will try to develop a toy model that, although approximate, can at least qualitatively describe the high concentration viscosity trends of \ce{FeCl_2} and \ce{MgCl_2}.

We introduce a second equilibrium constant, $K_2$, defined as the equilibrium constant for the reaction $ \ce{MCl^{+}} +  \ce{Cl^{-}} \rightleftharpoons \ce{MCl_2^{0}}$ according to:
\begin{equation}
    K_2 = \frac{ a_{ \ce{MCl_2^{0}}} }{ a_{ \ce{MCl^{+}} } \; a_{ \ce{Cl^-} } } = \frac{ \ce{[MCl_2^{0}]} }{ \ce{[MCl^{+}]} \ce{[Cl^{-}]} } \frac{\gamma_{ \ce{MCl_2^{0}} } } {\gamma_{ \ce{MCl^{+}} } \; \gamma_{ \ce{Cl^-} }},
    \label{def_K2}
\end{equation}
where symbols have the same meaning as in Eq.~\eqref{def_K1}.

In addition to relations for the equilibrium constants, we need a model to describe how the viscosity changes as the population of complexes changes.
We propose a simple and heuristic approach, and suggest the following equation to estimate the viscosity at $4$ m for \ce{MgCl2} and \ce{FeCl2}
\begin{equation}
    \eta (4 \ \mathrm{m} ) = \eta^{0} - \delta' \alpha' - \delta'' \alpha'' 
    \label{eta_as_thermometer}
\end{equation}
where $\eta^{0}$ is the viscosity (in mPa$\cdot$s) 
%obtained at $4$ m with the system 
of a solution with $100\%$ free cations.

The basis for this formula is the observation that the viscosity changes linearly with the number of monomers, in the absence of dimers, and also that the viscosity changes linearly (albeit with a different slope) with the number of dimers, in the absence of monomers.
% Therefore, Eq.~\eqref{eta_as_thermometer} produces an estimate of the viscosity of $\ce{FeCl_2}$ and \ce{MgCl_2} at $4$ m, given the percentage of monomers ($\alpha'$) and dimers ($\alpha''$).
The value of the slopes, can be obtained from the simulations of this work. 
We obtain values of $\delta'_{q=0.85 \ e}=0.020$ mPa$\cdot$s and $\delta'_{q=0.80 \ e}=0.010$ mPa$\cdot$s for the $q=0.85 \ e$ and $q=0.80 \ e$ models, respectively, from the slopes in Fig.~\ref{fig:visc_change_with_monomers_4m}. For $\delta''$, we obtain $\delta''_{q=0.85 \ e}=0.029$ mPa$\cdot$s and $\delta''_{q=0.8 \ e}=0.014$ mPa$\cdot$s (derived in Sec. 11 in the SI).

Secondly, we need the values of the association constants $K_1$ and $K_2$. We shall use the NIST value for $K_1$ of \ce{FeCl_2} [i.e., $\log(K_1)=-0.20)$]. 
% What value to use for $K_2$ ?
However, obtaining a value for $K_2$ is more cumbersome.
As shown in Sec.~11 of the SI, $\log K_2$ is often around $0.8$ units smaller than $\log K_1$ (see also Table 6.1 in \citet{burgess_book2}).
Therefore, we adopt the values of $\log K_2$ for \ce{FeCl_2} and \ce{MgCl2} listed in Table~\ref{tab:qualitative calculation}.
% as illustrated in Table \ref{tab:qualitative calculation}. 
% We assume a value of $-1.1$ for the $log K_1$ of $MgCl_2$.
% Similar to the case of \ce{FeCl2}, we assume that $log K_2 = -1.9$ for $MgCl_2$.
Note that we imposed a smaller value of $K_1$ for \ce{MgCl_2}, compared to that of \ce{FeCl_2}.
This was motivated by the observation that \ce{MgCl_2} is an archetypal strong 2:1 electrolyte that exhibits little complexation in concentrated solutions.~\cite{Friesen2019}

% details are in the main text, so I shortened the caption.
\begin{table}[hbt!]
\begin{threeparttable}
    \caption{Values of equilibrium constants\tnote{*} used for a qualitative calculation of the viscosity of \ce{FeCl_2} and \ce{MgCl_2} at $4$ m, estimated complex distributions (exemplified by $\alpha$, $\alpha'$, $\alpha''$), and the estimated viscosity, $\eta$, calculated from Eq.~\eqref{eta_as_thermometer}, using $\eta^0=5.13$ mPa$\cdot$s (the value of the viscosity obtained with the 0.8 charge model) for both solutions. 
    } 
    \label{tab:qualitative calculation}
    \small
    \begin{tabular}{lcccccc}
        \mytoprule
        System & $\log K_1$ & $\log K_2$ & $\alpha$ & $\alpha'$ & $\alpha''$ & $\eta$ (mPa$\cdot$s)\\
        \midrule
        \ce{FeCl_2} & $−0.20$ & $−1.0$ & $22$ & $56$ & $22$ & $4.46$ \\
        \ce{MgCl_2} & $−1.1$  & $−1.9$ & $64$ & $33$ & $3$ & $4.96$ \\
        \mybottomrule                             
    \end{tabular}
    \begin{tablenotes}\footnotesize
        \item[*]obtained from NIST, Lange's Handbook\cite{lange} or estimated using reasonable guesses; see also Sec. 11 of the SI.
    \end{tablenotes}
\end{threeparttable}
\end{table}
% and the results of this work for the 0.8 Madrid-2019 model for which the value of $\eta^0$ is $5.13$ mPa$\cdot$s for both solutions

Typically, the values of $K$ are determined from the concentrations at infinite dilution, where the activity coefficients tend towards one.
However, at finite concentrations, individual activity coefficients should be estimated, which is usually done using theoretical models, such as SIT\cite{sit_bronsted,sit_guggenheim,sit_ciavatta} (see also Sec.~\ref{subsec:inconsistencies}) or via simulations,\cite{pana_individual} using certain approximations.
Note that what is needed is not necessarily each individual activity coefficient, but the right hand side of Eq.~\eqref{def_K1} or Eq.~\eqref{def_K2}. 
  
It is true that this term is probably not equal to  one. Whatever the true value of this term is, one could consider this term to be effectively incorporated into the the value of $K_1$ and $K_2$, thus defining an effective equilibrium constant $K_1^{*}$ or $K_2^{*}$.
These are not true thermodynamic equilibrium constants but they are, instead, effective constants that are used with concentrations, and not with activities (they are usually denoted in the literature as stoichiometric equilibrium constants \cite{seawater_book}, and in contrast to the true equilibrium constants that depend only on temperature, these constants also depend on the media in which the reaction takes place). 
In this illustrative calculation, we assume that all values of $\gamma$ are unity and that Table ~\ref{tab:qualitative calculation} reports the true equilibrium constant, or that, equivalently, Table~\ref{tab:qualitative calculation}  reports the values of stoichiometric equilibrium constants $K_1^{*}$ and $K_2^{*}$ (with which one would use the concentrations and not activities). 
In any case, our approach to the problem is meant to be qualitative, rather than quantitative.
Moreover, this assumption yielded free cation populations which are in good agreement with the SIT model, when only monomers were considered, as shown in Table 
~\ref{free_cations_vmintec_4_0}.
% and that assuming $\gamma=1$ provided values of $\alpha$ in very good agreement with the predictions of the SIT activity model when only monomers were considered as was shown in Table 
% \ref{free_cations_vmintec_4_0}.

Thus, with the values of the equilibrium constants of Table~\ref{tab:qualitative calculation} and using Eq.~\eqref{eta_as_thermometer} to predict the viscosity and the results of the force field with the scaled charge $0.8$, we obtain a viscosity of $4.46$ mPa$\cdot$s for \ce{FeCl_2} (compared to $4.17$ mPa$\cdot$s from our experiments; see Table~\ref{jacobo}) and $4.96$ mPa$\cdot$s for \ce{MgCl_2} (compared to the experimental value of $4.73$ mPa$\cdot$s; see also Sec.~2 in the SI), respectively.
Therefore, agreement with experiments is very good, within $5\%$, which is also consistent with our simulation results. The populations of free ions, monomers and dimers is presented in Table~\ref{tab:qualitative calculation}. 
This simple toy model is consistent with the existence of monomers and dimers in \ce{FeCl_2} solutions, as shown in recent experiments.\cite{Zhao2001, Luin2022}

% Are we stating that Table \ref{tab:qualitative calculation} corresponds to the true populations of complexes that really exists in either $FeCl_2$ or $MgCl_2$ ? The
% answer to this question is negative. 
However, we emphasize that we are not claiming that the complex distributions obtained from this calculation correspond to that found in  real solutions of \ce{FeCl2} and \ce{MgCl2}.
We are merely showing how a certain population of complexes would be compatible with the experimental values of the viscosities.  
Note that there could be several complex distributions that are consistent with the experimental values. 
The solid orange line in Fig.~\ref{fig:visc_change_with_monomers_4m} (which assumes the absence of dimers) is also able to describe the viscosities of both solutions. 
% moved to the end
% Other properties, such as osmotic coefficients,~\cite{good_bad_hidden} freezing point depression,~\cite{freezing-cintia} or electrical conductivities~\cite{blazquez2023computation} could  be useful in evaluating different speciation possibilities, since we expect that they are likely to be sensitive to the presence of complexes. 
% In fact it is likely that these properties will also be quite sensitive to the presence of complexes and that should investigated in future work. 
However, in every case examined in this work, there is more association in \ce{FeCl2} than in \ce{MgCl2}, regardless of the exact proportions of complexes estimated in each solution.
% What is common between the orange line of Fig.4 and Table \ref{tab:qualitative calculation} is the conclusion that more association is needed in \ce{FeCl_2} compared to \ce{MgCl_2} in order to describe the experimental values of the viscosity.
In the future, properties such as osmotic coefficients,~\cite{good_bad_hidden} freezing point depression,~\cite{freezing-cintia} or electrical conductivities~\cite{blazquez2023computation} could be useful for evaluating different speciation possibilities, since we expect them to also be sensitive to the presence of complexes. 

\subsection{Limitations of empirical force-fields}
\label{subsec:limitations}
Most of the ions described up to this point by the Madrid-2019 force-field have an electronic configuration close to that of noble gases (halogen, alkaline, alkaline-earth), and for those the model has been relatively successful.
Modeling \ce{Fe^{2+}} with this force-field is the first foray into transition metals which has revealed various challenges and nuances tied to complex formation.
% Iron (\ce{Fe^{2+}}) represents the first foray into transition metals which has revealed different challenges and nuances, tied to complex formation.
An interesting future challenge will be a study of solvated transition metal ions with larger charge, such as
\ce{Fe^{3+}}. 
Recent simulations have shown abrupt switching between two different 
solvation shell structures, with lifetimes of the order of nanoseconds.\cite{Goswami2024}

However, even with an empirical force-field, it is possible to increase the population of \ce{FeCl^{+}} monomers with respect to \ce{MgCl^{+}}.
This could be accomplished by having different LJ parameters for the \ce{Fe-Cl} and \ce{Mg-Cl} interactions.
For instance, decreasing the value of $\sigma$ (the size parameter in the LJ potential), in the first case, with respect to the second, would increase the strength of the \ce{Mg-Cl} interaction and would consequently increase the population of monomers.
This approach was used by \citet{DuboueDijon2017} to increase the number of CIPs in \ce{ZnCl_2}.
Although this methodology is interesting, we believe that it is more fruitful to recognize
%, at this point, 
the limits of empirical force fields to quantitatively describe the energy between a transition metal with several  $3 \ d$ valence electrons and a ligand, and to 
%openly 
acknowledge that only 
%a quantum chemistry treatment 
an electronic structure calculation
can describe this interaction in a quantitative way.
Since the required size of a simulated system is large and the computational effort of electronic structure calculations scales rapidly with the number of electrons, a practical approach could involve a hybrid simulation method, such as QM/MM (quantum mechanics/molecular mechanics).\cite{Kirchhoff2021}

%%%%%%%%%%%%%%%%%%%%%%%%%%%%%%%%%%%%%%%%%%%%%%%%%%%%%%%%%%%%%%%%%%%%%
%% Conclusions
%%%%%%%%%%%%%%%%%%%%%%%%%%%%%%%%%%%%%%%%%%%%%%%%%%%%%%%%%%%%%%%%%%%%%
\section{Conclusions}

The literature contains numerous conflicting reports on complex speciation and the extent of complexation in concentrated \ce{FeCl2} and \ce{MgCl2} solutions.
In general, using experiments, theory or simulations to estimate equilibrium complex populations in concentrated solutions is a non-trivial task.
However, knowledge of speciation is crucial to understand the behaviour of solutions.

Modelling a transition metal, such as \ce{Fe^2+}, with an empirical force-field is complicated by electronic structure effects (for instance, $3d^6$ configuration).
In light of this, the extension of the Madrid-2019 force-field to \ce{Fe^2+} may be deceptively simple, but is a research outcome in its own right. 

We have introduced a methodology to incorporate complexes in our simulations which is similar, in spirit, to the seeding method for nucleation, \cite{espinosa_seeding_2016, bai2006calculation,goswami2020seeding} wherein prefabricated critical clusters are introduced.
Both methods are characterized by two different timescales: a significantly large one (corresponding to the attainment of the equilibrium complex population in concentrated solutions, in this case, or to the emergence of a critical nucleus in nucleation), and another relatively short timescale of interest (corresponding to transport property calculations as in this work, or to the evolution of a critical cluster in nucleation).
Inserting complex populations in simulations of concentrated solutions \emph{a priori} enables us to bypass the computationally prohibitive time required to reach these equilibrium populations.
Note that one could obtain information about plausible complex populations from various sources, such as experiments, first principles, and thermodynamics, and subsequently perform simulations using any standard force-field.

We have presented an argument for a strong link between complex distributions and the viscosity, using computer simulations (wherein input complex populations were varied) and with new experimental data for \ce{FeCl2} at high concentration.
Although the methods described in this work cannot quantitatively estimate complex populations, our results provide qualitative information about speciation and show that \ce{FeCl2} has a higher propensity of complexation than \ce{MgCl2}. 
This could be particularly valuable to obtain clarity about complex speciation when experimental data is elusive or inconclusive.

% moved to last paragraph
% We should mention that experimental results using  
% NDIS (Neutron Diffraction with Isotopic Substitution\cite{review_ndis_1,review_ndis_2,review_ndis_3}) using two isotopes of $Cl$ in "null" water (a mixture of $H_2O$ and $D_2O$ ) would be extremely useful to clarify the existence and amount of $M-Cl$ complexes. In fact that was done recently\cite{good_bad_hidden} to clarify the number of CIP in $CaCl_2$. As H do not contribute to the scattering in null water and the $Cl-O_w$ distance is around 3.14 \AA whereas the $Cl-Mg$ would be located around 2.33 \AA the two peaks will not overlap, and the experimental results will determine without ambiguity the existence of complexes and the amount of them at high concentrations. 

This idea is not as novel as it may first seem.
\citet{hertz} previously reported experimental viscosity measurements for \ce{MgCl2} and \ce{ZnCl2}, observing that the two systems exhibit very similar viscosities at low concentrations but diverge significantly at higher concentrations.
% Several years previously, the viscosity of \ce{MgCl_2} and \ce{ZnCl_2} was measured experimentally by \citet{hertz}
% They also realized that, at low concentrations, the viscosities of both systems were almost the same, but found significant differences at high concentrations.  
The authors speculated that the formation of complexes caused the lower viscosity of \ce{ZnCl_2}. 
While one cannot change the number of complexes at will in experiments, this can be easily achieved in simulations.
% While in experiments one cannot change the number of complexes at will, this can be easily achieved in simulations. 
In this work, we have demonstrated how complexes can affect the viscosity of solutions at high concentrations, thus confirming the intuition of \citet{hertz}
Nevertheless, further experimental work is required to quantitatively clarify the amount of complexes in concentrated solutions, and NDIS (Neutron Diffraction with Isotopic Substitution\cite{review_ndis_1,review_ndis_2,review_ndis_3}), using chloride isotopes in “null” water (a mixture of \ce{H_2O} and \ce{D_2O}),\cite{good_bad_hidden} could be particularly informative, especially since the chloride-oxygen chloride-cation peaks would not overlap (expected to be around $3.14$ \AA ~and $2.33$ \AA, respectively).
We hope that future research will be continued in this direction, which was first envisaged by \citet{hertz} more than $40$ years ago in this journal. 
\begin{suppinfo}

The authors confirm that the data supporting the findings of this study are available within the article and/or its supplementary materials.
A workflow for creating configurations, running LAMMPS simulations and calculating the viscosity can be found at \url{https://github.com/amritagos/electrolyte_workflow}.
The supporting information contains a description of the simulation methodology for molecular dynamics (MD) simulations, 
details about the experimental viscosity of \ce{MgCl_2} used in this work,
an expanded discussion of the justifications for using the force-field parameters of \ce{Mg^2+} for \ce{Fe^2+}, 
rationale for the cation-chloride distances chosen to create complexes in simulations (by freezing the distance) and further justifications for this approach,
an analysis of solvation shell structures showing that freezing cation-chloride distances does not change the solvation shell,
theoretically expected viscosity trends of \ce{MgCl2} and \ce{FeCl2} and details about approximating electrolyte solutions as LJ fluids in order to estimate expected viscosity,
predictions of complex speciation from Visual MINTEQ,
densities calculated from simulations with and without complexes,
a discussion of the slight overestimation of the viscosity by the Madrid-2019 force-field,
the effects of complex populations on transport properties,
additional details about a back-of-the-envelope calculation for estimating viscosity from Eq.~\eqref{eta_as_thermometer},
and details of how Jones-Doles $B$ coefficients were estimated using the force-field.
\end{suppinfo}

%%%%%%%%%%%%%%%%%%%%%%%%%%%%%%%%%%%%%%%%%%%%%%%%%%%%%%%%%%%%%%%%%%%%%
%% The "Acknowledgement" section can be given in all manuscript
%% classes.  This should be given within the "acknowledgement"
%% environment, which will make the correct section or running title.
%%%%%%%%%%%%%%%%%%%%%%%%%%%%%%%%%%%%%%%%%%%%%%%%%%%%%%%%%%%%%%%%%%%%%
\begin{acknowledgement}
This work was funded by the Icelandic Research Fund (grants 228615-051 and 2410644-051),
and by the Spanish Ministry of Science and Innovation (grants PID2022-136919NB-C31,
 PID2023-151751NB-I00 and PID2023-147148NB-I00).
The calculations were performed using compute resources provided by the Icelandic Research Electronic Infrastructure (IREI).
L.F.S. acknowledges an FPU predoctoral Grant No. FPU22/02900.
A.G. is grateful to Moritz Sallermann, Rohit Goswami, Alejandro Peña-Torres and Elvar \"Orn J\'{o}nsson for fruitful discussions.
\end{acknowledgement}

\putbib  % prints the main-text bibliography here

\end{bibunit}

%%%%%%%%%%%%%%%%%%%%%%%%%%%%%%%%%%%%%%%%%%%%%%%%%%%%%%%%%%%%%%%%%%%%%
%% SUPPORTING INFORMATION
%%%%%%%%%%%%%%%%%%%%%%%%%%%%%%%%%%%%%%%%%%%%%%%%%%%%%%%%%%%%%%%%%%%%%

\begin{bibunit}
\renewcommand{\bibnumfmt}[1]{S#1.}
\renewcommand{\citenumfont}[1]{S#1}
\setcounter{NAT@ctr}{0}

\clearpage
\onecolumn

\begin{center}
{\Large \textbf{Supporting Information}}\\[1em]
{\large Viscosity as a Smoking Gun for Complex Formation in Solution: \ce{Fe^{2+}} and \ce{Mg^{2+}} Chlorides as Examples}\\[1em]
Amrita Goswami,
Samuel Blazquez,
Lucía Fern{\'a}ndez-Sedano Vázquez,
Eva Gonz{\'a}lez Noya,
Hannes J\'{o}nsson,
Jacobo Troncoso,
Carlos Vega
\end{center}

\vspace{1cm}

% remove main-paper TOC entries
\makeatletter
\immediate\write\@auxout{\string\@writefile{toc}{}}
\makeatother

\begingroup
\renewcommand{\contentsname}{}
\tableofcontents
\endgroup
\clearpage

% S-numbering...
\setcounter{section}{0}
\setcounter{figure}{0}
\setcounter{table}{0}
\setcounter{equation}{0}
\renewcommand{\thesection}{S\arabic{section}}
\renewcommand{\thefigure}{S\arabic{figure}}
\renewcommand{\thetable}{S\arabic{table}}
\renewcommand{\theequation}{S\arabic{equation}}

\changelocaltocdepth{3}

%%%%%%%%%%%%%%%%%%%%%%%%%%%%%%%%%%%%%%%%%%%%%%%%%%%%%%%%%%%%%%%%%%%%%
%% Start the main part of the supplementary here.
%%%%%%%%%%%%%%%%%%%%%%%%%%%%%%%%%%%%%%%%%%%%%%%%%%%%%%%%%%%%%%%%%%%%%
\section{Simulation methods}
\label{sec:simulation_method}

In this work, we have extended the Madrid-2019 force-field~\cite{zeron2019force} to the Ferrous ion, \ce{Fe^2+}, using the same parameters as for \ce{Mg^{2+}}.
The Madrid-2019 force-field combines the TIP4P/2005 model of water\cite{tip4p/2005} with scaled charges for the ions.
The proposed force-field parameters are collected in Table 1 in the main text.
Parameters for \ce{Mg^{2+}}, \ce{Cl^{-}} can be found in our previous work~\cite{zeron2019force} and those of perchlorate from our recent work.\cite{blazquez2025extending} 
Note that the canonical Madrid-2019 force-field~\cite{zeron2019force} uses a scaled charge of $0.85$ \textit{e}. 
It is well-known that the Madrid-2019 force-field tends to overestimate the viscosities of electrolytes in water 
(even for \ce{NaCl}, \ce{KCl} where contact ion pairs, but no complexes, are formed), as described further in Sec~\ref{sec:madrid_visc_overestimation}. 
It has been proposed that lowering the charge to $0.8$ reduces many of the discrepancies with experiments 
(and with $0.75$ it is even possible to reproduce the experimental results).\cite{blazquez2023scaled} 
For this reason, we have also performed analyses using a scaled charge of $0.8$. 
The change of the charge of the ions requires some adjustment of the Lennard-Jones (LJ) parameters (simply replacing the charge and keeping the LJ parameters optimized for the $q=0.85$ \textit{e} model is not a good option\cite{blazquez2023scaled}).
For the model with $q=0.8$ \textit{e}, the value of  $\sigma$ for the \ce{Cl-$\mathrm{O_w}$} was taken from our previous work Ref.\cite{blazquez2023scaled} 
and the value of LJ $\sigma$ for the \ce{Fe^{2+}-$\mathrm{O_w}$} or \ce{Mg^{2+}-$\mathrm{O_w}$} was optimized in this work, obtaining $\sigma=1.85$ \AA, as shown in Table 1 in the main text.

Regarding \ce{FeCl_2} simulations, it is important to consider the possibility of \ce{Fe^{2+}} undergoing a certain amount of hydrolysis in water. 
However, in this work, for the highest concentration ($4$ m), the pH is about $5$. 
Therefore, the concentration of \ce{H3O^+} and \ce{Fe(OH)^+} is quite small (of the order of about $10^{-5}$ m) and can be neglected when performing simulations to describe this system.

We have used the molality scale, \textit{m}, for concentrations (mol of salt per kg of water) as is typically done in experimental studies, since the concentration then does not depend on the volume of the system. 
Furthermore, in simulations we typically used $555$ molecules of water to determine the densities of the aqueous ionic solutions (and multiples of the same for viscosity simulations). 
This is very convenient, since then $1$ m or $2$ m solutions are obtained with $10$ or $20$ molecules of salt. 

\subsection{Simulation details}
\label{subsec:sim_details}

Simulations were performed using the GROMACS (version 2025.2)~\cite{Abraham2015} and LAMMPS (29 Aug 2024, Update~1)~\cite{Thompson2022} packages.
Specific simulation details for GROMACS and LAMMPS are provided in Sec.~\ref{subsec:sim_details_gromacs} and Sec.~\ref{subsec:sim_details_lammps}, respectively.

Densities were computed from NpT simulations, with $555$ water molecules, run for $50$ ns each. 
Prior to production runs, the simulations were equilibrated for $10$ ns.

For the viscosity calculations, each independent configuration contained $4440$ water molecules, within a box size corresponding to the averaged density for each concentration.
Note that independent configurations were created using the methodology described in Sec. ~\ref{sec:config_generation}.
Viscosities were computed from $8-15$ independent simulation runs in the NVT ensemble, using both LAMMPS and GROMACS, which were subsequently averaged (over a total of $16-30$ independent simulations). Each independent simulation was run for $40$-$50$ ns (following a $4$ ns simulation of simulated annealing and subsequently a $20$ ns period of equilibration). 

The diffusion coefficient was calculated from unwrapped coordinate trajectories generated during the viscosity calculations in the NVT ensemble. 
The Yeh-Hummer correction~\cite{Yeh2004} was applied to the diffusion coefficients to account for finite-size effects. 

A workflow for creating configurations, running LAMMPS simulations and calculating the viscosity can be found at \url{https://github.com/amritagos/electrolyte_workflow}. 
Snapshots and visuals were created using \texttt{solvis} (\url{https://github.com/amritagos/solvis}). 

\subsubsection{GROMACS}
\label{subsec:sim_details_gromacs}
In GROMACS, simulations were performed in the NpT ensemble with a Nose-Hoover thermostat~\cite{Evans1985} and the barostat from
Parrinello-Rahman.~\cite{Parrinello1981,Nose1983} 
The time step was of $2$ fs.
Ewald sums (PME) were used for the long-range electrostatic interactions in GROMACS.
A long-range correction to the LJ part of the potential was added.
The potential was truncated at 10 {\AA}. 
The shape of the water molecules was constrained using the LINCS algorithm~\cite{hess1997lincs}. 
In the case of complexes, the cation-anion distance was also frozen using LINCS (see also Sec.\ref{sec:chosen_distances} and Sec.\ref{sec:validation}).
\subsubsection{LAMMPS}
\label{subsec:sim_details_lammps}
Analogous simulations in LAMMPS were performed using the Nose-Hoover thermostat~\cite{Evans1985} and Nose-Hoover barostat,~\cite{Shinoda2004} with a thermostat and barostat coupling constant of $200$ fs and $5$ ps, respectively.
Likewise, the timestep used was $2$ fs.
Long-range electrostatics were computed using particle-particle particle-mesh (PPPM).~\cite{Hockney1988}
Similar to the protocol in GROMACS, long-range corrections to the LJ terms were added, and a cutoff of 10 {\AA} was used.
The SHAKE algorithm~\cite{Ryckaert1977} was used to constrain the shape of water molecules and to constrain the cation-anion distances in complexes (see also Sec.\ref{sec:chosen_distances} and Sec.\ref{sec:validation}).

\subsection{Configuration generation with complexes and energy minimization}
\label{sec:config_generation}

As part of our new simulation strategy of studying transport properties at a static predetermined complex population (see Sec.\ref{sec:chosen_distances} and Sec.\ref{sec:validation} for justification and validation), we perform simulations in which we artificially maintain the desired complex population. 
This is accomplished by constraining the cation-anion distance in simulations of \ce{FeCl_2} and \ce{MgCl_2}.
Concerning configuration generation, usually it is trivial to create initial configurations using tools such as PACKMOL~\cite{Martinez2009} or Moltemplate.~\cite{Jewett2021} 
However, especially for concentrated systems, and in the situations where we insert complexes, configuration creation can be a slightly more involved process.
In this work, a \emph{monomer} refers to a complex formed by one cation, coordinated with one \ce{Cl^{-}} anion and some additional molecules of water in the first solvation shell, while a \emph{dimer} refers to a complex wherein there are two \ce{Cl^{-}} anions in the trans position of the octahedron, and some additional coordinating water molecules (see Fig.2 in the main text).
In the case of a "monomer" or "dimer" complex, a unit consisting of a bonded metal ion and chloride, or a "triatomic" unit consisting of two chlorides and one metal ion was created. 
Each species was randomly distributed in the simulation box, keeping a tolerance of 3.0 {\AA} between complexes.
A \texttt{Python} program was written for this purpose which uses PACKMOL, Moltemplate and Atomic Simulation Environment (ASE)~\cite{HjorthLarsen2017}, under the hood; see \url{https://github.com/amritagos/confi}.
Certain typical file formats in LAMMPS and GROMACS can also be converted using \texttt{confi}.

Configurations were created in the manner described above in order to facilitate subsequent energy minimization. 
Energy minimization with the aforementioned complexes was performed by keeping the complexes frozen and only allowing the water molecules to move. However, in the case of free cations and anions, all atoms were allowed to move. 
The energy was minimized in two stages-- first with $1000$ steps of steepest descent, followed by a maximum of $10000$ steps of the conjugate gradient algorithm, with a step size of $0.001$ fs. 

\subsection{Viscosity calculations from simulations}
\label{sec:viscosity}

Using the Green-Kubo formalism, the viscosity is obtained from the expression:
\begin{equation}
    \label{shear_GK}
    \eta = \frac {V}{k_BT} \int_{0}^{\infty} 
    \langle  P_{\lambda\beta} (\tau)\; P_{\lambda\beta}(0) \rangle\; d\tau,
\end{equation}
with $k_B$ the Boltzmann constant, $V$ is the volume, $T$ is the temperature and $P_{\lambda\beta}$ are the non diagonal components of the pressure tensor (i.e., $\lambda \neq \beta$) which are given as:

\begin{equation}
    \label{pressure_tensor}
    \begin{aligned}
        P_{\lambda\beta} (\tau) \,=\, &\frac{1}{V} \sum_{i=1}^{N}  m_i v_{\lambda,i}(\tau) v_{\beta,i}(\tau)\\ 
        + &\frac{1}{V} \sum_{i=1}^{N}  r_{\lambda,i}(\tau) f_{\beta,i}(\tau)\\ 
        \,=\, &P_{\lambda\beta}^{K} + P_{\lambda\beta}^{U},
    \end{aligned}
\end{equation}

where $v_{\lambda,i}$, $r_{\lambda,i}$  and $f_{\lambda,i}$ denote  the  $\lambda$ component of the velocity, position and force acting on particle i respectively. The first term of Eq.~\eqref{pressure_tensor} is a kinetic term that depends on the masses of the particles of the system, and the second term is a virial term that only depends on positions and forces, but not on masses. 
The sum in Eq.~\eqref{pressure_tensor} is over the $N$ particles of the system (water and ions, in our case).
The three non diagonal components and the terms $\frac{1}{2}$ ($P_{xx}-P_{yy}$) and  $\frac{1}{2}$  ($P_{yy}-P_{zz}$) were used to get the correlation function of the pressure tensor as in the work of \citet{abascal_viscosity}
The pressure tensor can be divided into a kinetic contribution K, and a potential energy contribution U as shown in Eq.~\eqref{pressure_tensor}.
Therefore when using the Green-Kubo formula one has a KK contribution, a KU contribution and a UU contribution. 

\section{Experimental viscosities of \ce{MgCl_2}}
\label{sec:experimental_visc_MgCl2}

\begin{figure}[t]
  \includegraphics[width=0.6\linewidth]{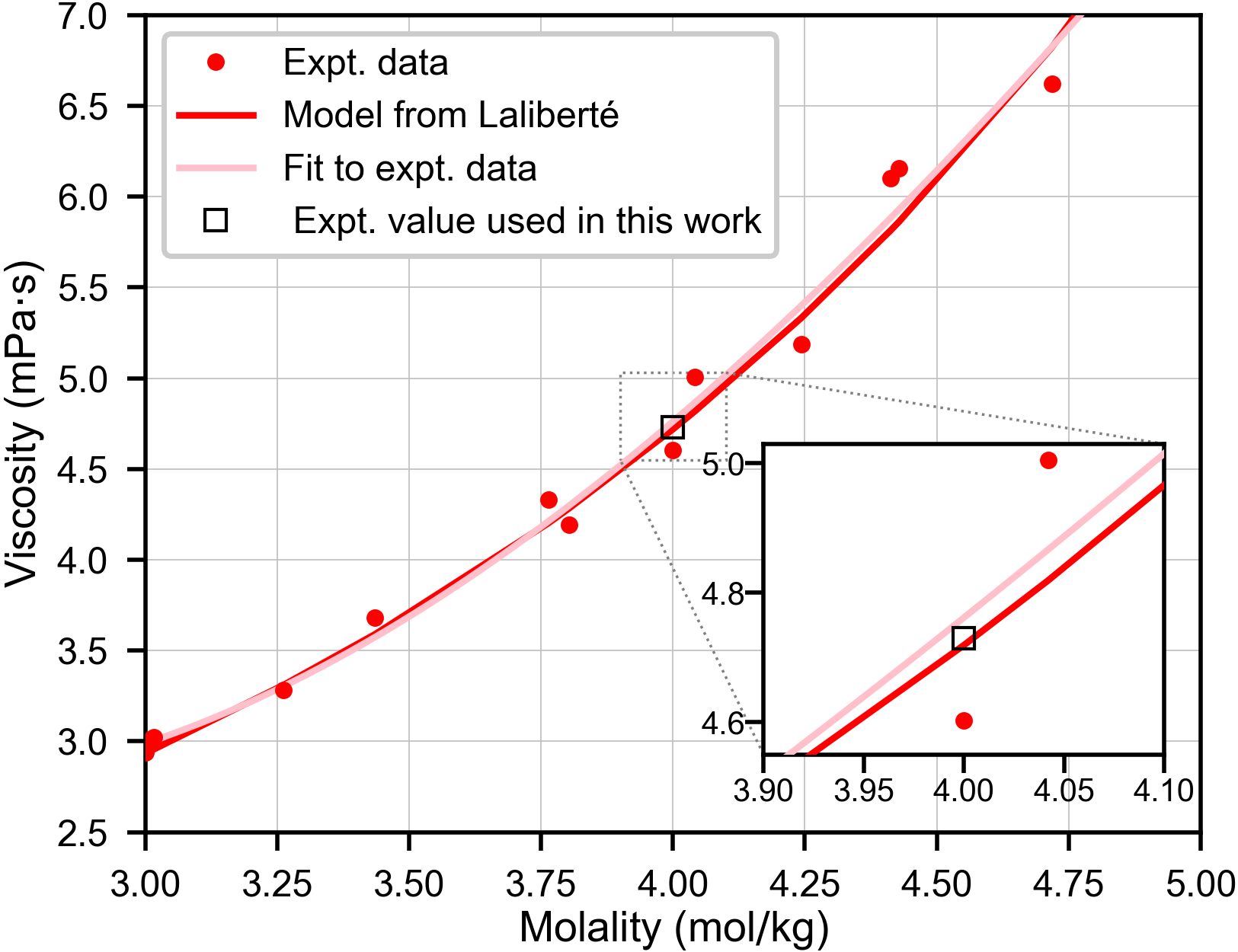}
  \caption{\label{fig:mgcl2_expt_chaos} Viscosities of \ce{MgCl_2} from experiments\cite{laliberte2007model} (filled red circles), model fitted to experimental data points\cite{laliberte2007model} (solid red line), along with a fit to experimental data points in the interval $3$ m to $5$ m (solid pink line). The experimental value of the viscosity used in this work at $4$ m is denoted by an open square. The inset shows a zoomed-in view close to $4$ m. 
  }
\end{figure}

As shown in Fig.~\ref{fig:mgcl2_expt_chaos}, individual experimental measurements (red circles) are scarce and rather scattered. Therefore, \citet{laliberte2007model} has proposed a model that fits to the existing experimental data (solid red line in Fig.~\ref{fig:mgcl2_expt_chaos}).
In this work, we have performed a detailed study at $4$ m.
Therefore, we have also fitted to the experimental data in the range $3$-$5$ m. 
Using the model from \citet{laliberte2007model} and our own fit, we settle on a viscosity of $4.73$ mPa$\cdot$s for \ce{MgCl_2} at $4$ m.
Subsequently, we have used this value in this work as the experimental value of \ce{MgCl2}. 

\section{Justification for the \ce{Fe^2+} force-field parameters}

\begin{figure*}
  \includegraphics[width=\linewidth]{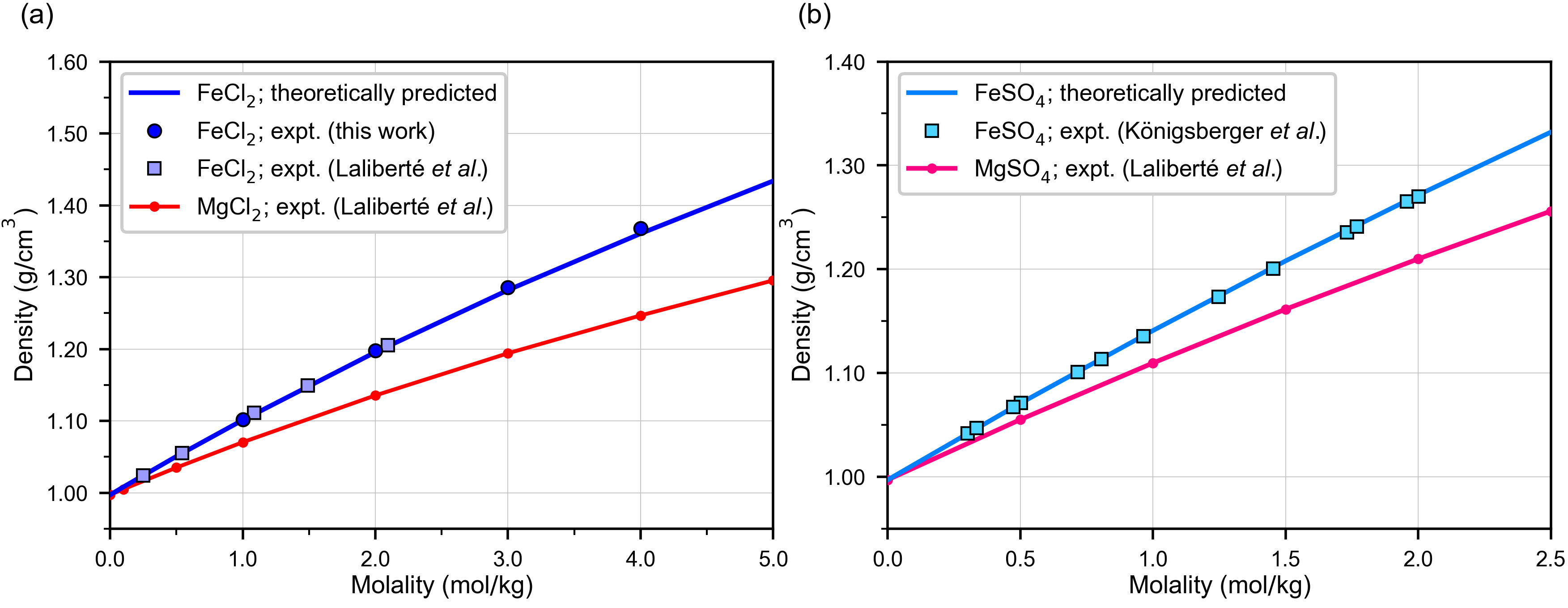}
  \caption{\label{fig:equal_volumes} (a) Experimental densities of \ce{FeCl2} (light-blue symbols and blue symbols for results from~\citet{Laliberte2004} and results from this work, respectively), as well as experimental densities of \ce{MgCl2} (red line).
  Assuming equal volumes for the \ce{FeCl2} and \ce{MgCl2} solutions, we obtain the solid blue line as the predicted density of \ce{FeCl2}. 
  (b) Densities of \ce{FeSO_4} and \ce{MgSO_4} from experiments, and those scaled by assuming equal volumes. 
  Densities of \ce{FeSO_4} were taken from~\citet{Koenigsberger2008}, and those of \ce{MgSO_4} from \citet{Laliberte2004}.}
\end{figure*}

Fig.~\ref{fig:equal_volumes}(a) presents the densities of \ce{FeCl_2} and \ce{MgCl_2} obtained from experiments, including the experimental results of this work (see also Table 2 of the main text).
As can be gleaned from Fig.~\ref{fig:equal_volumes}(a), the mass densities of \ce{FeCl_2} and \ce{MgCl2} are clearly different. 

Using the approximation of equal number densities (Eq.~1 in the main text), we can theoretically estimate the mass density of \ce{FeCl2}, using the experimentally calculated mass densities of \ce{MgCl2}.
Fig.~\ref{fig:equal_volumes}(a) shows that the theoretically predicted mass densities of \ce{FeCl2} (assuming that \ce{MgCl2} and \ce{FeCl2} have the same number density) agree well with the densities from experiments.
We show that a similar approximation can be made for \ce{FeSO_4} and \ce{MgSO_4} solutions in Fig.~\ref{fig:equal_volumes}(b).

This is a promising sign that the same force-field parameters as \ce{Mg^{2+}} could be used for \ce{Fe^{2+}}.

\begin{table}[hbt!]
  \centering
  \captionsetup{justification=centering}
  \caption{
      Metal -- $\mathrm{O_w}$ distances, $d_{\ce{M-$\mathrm{O_w}$}}$ (in \AA), and hydration free energies (in kcal/mol).\tnote{*}
    }
    \label{distances_metal_OW}
  \begin{threeparttable}
    \small
    \begin{tabular}{l c r}
      \mytoprule
      & $d_{\ce{M-$\mathrm{O_w}$}}$ & HFE \\
      \midrule[0.05em]
      \ce{Mg^{2+}} -- $\mathrm{O_w}$ & $2.11 \pm 0.01^{\mathrm{b}}$ & $-437.4^{\mathrm{c}}$ \\[0.5em]
      \midrule[0.01em]
      \multirow{2}{*}{\ce{Fe^{2+}} -- $\mathrm{O_w}$}
        & $2.095^{\mathrm{a}}$ & \multirow{2}{*}{$-439.8^{\mathrm{c}}$} \\
        & $2.09 \pm 0.04^{\mathrm{b}}$ & \\
      \mybottomrule
    \end{tabular}
    \begin{tablenotes}\footnotesize
      \item[*] a) From \citet{fecl2_exp_xray}
      b) Collected in \citet{Ohtaki1993, marcus1988ionic}
      c) From \citet{marcus1991thermodynamics}
    \end{tablenotes}
  \end{threeparttable}
\end{table}

% \begin{table}[hbt!]
%   \caption{
%   Metal -- $\mathrm{O_w}$ distances, $d_{\ce{M-$\mathrm{O_w}$}}$ (in \AA ), and the hydration free energy from experimental results.
%   a) Results from \citet{fecl2_exp_xray}.
%   b) Results from ~\citet{Li2020}
%   The hydration free energies (HFE), in kcal/mol, are also shown.} 
%   \label{distances_metal_OW}
%   \small
%   \begin{tabular}{l c r  }
%       \mytoprule
%       & $d_{\ce{M-$\mathrm{O_w}$}}$ & HFE \\
%       \midrule[0.05em]
%       \ce{Mg^{2+}} -- $\mathrm{O_w}$  &  $2.11 \pm 0.01^{b}$  &  $-437.4^{b}$ \\[0.5em]
%       \midrule[0.01em]
%       \multirow{2}{*}{\ce{Fe^{2+}} -- $\mathrm{O_w}$} & $2.095^{a}$ & \multirow{2}{*}{$-439.8^{b}$ }  \\
%       & $2.09 \pm 0.04^{b}$ & \\
%       \mybottomrule
%   \end{tabular}
% \end{table}

We believe that this is further justified by the fact that the experimentally measured cation-oxygen distance\cite{Persson2010} is similar for \ce{Mg^{2+}} and \ce{Fe^{2+}}, as shown in Table~\ref{distances_metal_OW}.
Table~\ref{distances_metal_OW} also presents the hydration free energies, which are almost identical for both ions.
Therefore, we surmise that the cation-water interactions for both cations are similar, both in terms of the distance and magnitude of the interaction energy. 
Following on this premise, we conclude that a suitable force-field for \ce{Mg^{2+}} should work reasonably well for \ce{Fe^{2+}}.

Note that when the same force-field is used to describe \ce{Mg^{2+}} and \ce{Fe^{2+}}, despite having different masses,
classical simulations should necessarily yield the same thermodynamic properties such as the hydration free energy, radial distribution functions, surface tension, activity coefficients and freezing point depression.
This is a consequence of using classical statistical thermodynamics.
However, transport properties are affected by the mass and are, therefore, expected to be different.

\section{Justification for chosen cation-chloride distances and for rigid bonds}
\label{sec:chosen_distances}

\begin{table}[hbt!]
  \caption{Bond length distances, $d_{\mathrm{M-Cl}}$, (in \AA) in the metal ion chloride (\ce{M-Cl}) complexes. 
  % The experimental value for \ce{FeCl2} has taken from ~\citet{fecl2_exp_xray}
  % The \ce{Mg-Cl} distance is from a simulation study~\cite{Zapalowski2001} and from an experimental study of the crystal \ce{MgCl2(H2O)_4}, which forms octahedral complexes of \ce{Mg} with four water and two chloride anions in \textit{trans} position (see \citet{Schmidt2011}). The distance for the \ce{Fe^{3+}-Cl} has been taken from the crystalline form of the compound \ce{FeCl_3 (H2O)6},\cite{Lind1967} which is formed by \ce{Fe^{3+}} coordinated by four water and two chlorides in the \emph{trans} position.
  \tnote{*}} 
  \label{distances_complexes}
  \begin{threeparttable}
    \small
  \begin{tabular}{l c c c c }
      \mytoprule
      System &  $d_{\mathrm{M-Cl}}$    \\
      \midrule
      \ce{Mg^{2+}}-- Cl  &  $2.30^{\mathrm{a}}$, $2.55^{\mathrm{b}}$       &   &     &   \\ 
      % \ce{Mg^{2+}}-- Cl  &  2.55                  \\
      \ce{Fe^{2+}}-- Cl  &  $2.33^{\mathrm{c}}$   &   &     &   \\
      %$Zn^{2+}-Cl$  &  2.29   &   &     &     \\
      \ce{Fe^{3+}}-- Cl  &  $2.30^{\mathrm{d}}$    &   &     &      \\
      \mybottomrule
  \end{tabular}
\begin{tablenotes}\footnotesize
      \item[*] a) From a simulation study by ~\citet{Zapalowski2001}
      b) From an experimental study of \ce{MgCl2(H2O)_4} by~\citet{Schmidt2011}
      c) From~\citet{fecl2_exp_xray}
      d) From~\citet{Lind1967}
    \end{tablenotes}
  \end{threeparttable}
\end{table}

It has been shown that when the metal–water oxygen distance (M--O$_{\mathrm{w}}$) is similar for two cations, their corresponding metal–chloride (M--Cl) distances are also comparable. For example, \citet{kuppuraj2009factors} demonstrated this relationship for the divalent cations \ce{Co^2+} and \ce{Fe^2+}.

Metal–chloride distances for a range of divalent metal ions are reported in Refs.~\citep{Wei2023, Wrobel2020} and are summarized in Table~\ref{distances_complexes}. 
These data show that the M--Cl bond lengths for \ce{Fe^2+} and \ce{Mg^2+} fall within a narrow range.

On this basis, we adopt the same metal–chloride bond distance for both the Fe--Cl and Mg--Cl complexes, setting M--Cl = $2.33$~\AA.
Note that, in our new simulation approach, cation-anion distances are constrained to the same value using either the SHAKE~\cite{Ryckaert1977} or LINCS~\cite{hess1997lincs} algorithms.

We justify this new approach of freezing ion pairs in the following.
The lifetime of the cation-anion pair (complexes) in \ce{FeCl_2} and \ce{MgCl_2} is of the order of microseconds.~\cite{Shi2023}
Note that this means that a chloride ion cannot leave the \ce{Fe^2+} or \ce{Mg^2+} cation within our simulation time scales.
However, insofar as the viscosity is concerned, a lifetime greater than $\approx 100$ ps could be effectively considered "intransient" or "stable".
This is because the  Green-Kubo integral reaches a plateau, and that indicates that the integrand, which gives the correlation function of the non-diagonal pressure tensor components, goes to zero, as can be seen from the inset of Fig. 1 in the main text.
Thus, if the residence time of a cation and an anion is less than $100$ ps (which is the case for aqueous \ce{NaCl}, \ce{KCl} or \ce{RbCl}), then it is a labile contact ion pair, and it is more useful to consider it as a fluctuation within the configurational space.

As touched upon previously, the stability of long-lived complexes formed in \ce{FeCl_2} and \ce{MgCl_2} implies that reaching the equilibrium chemical composition can take of the order of microseconds to seconds (especially if the empirical force-field, or even an \emph{ab-initio} simulation, mimics the experimental relaxation times, in a realistic way, which is desirable).
When using the Madrid-2019 force-field, the formation of complexes is not observed even after $200$ ns runs
(i.e., the cation and anion were not in contact, in both the initial and in the final configurations).

\section{Validation of simulation protocol with rigid cation-chloride bonds}
\label{sec:validation}

\begin{figure}[ht]
    \centering
    \includegraphics[width=0.6\linewidth]{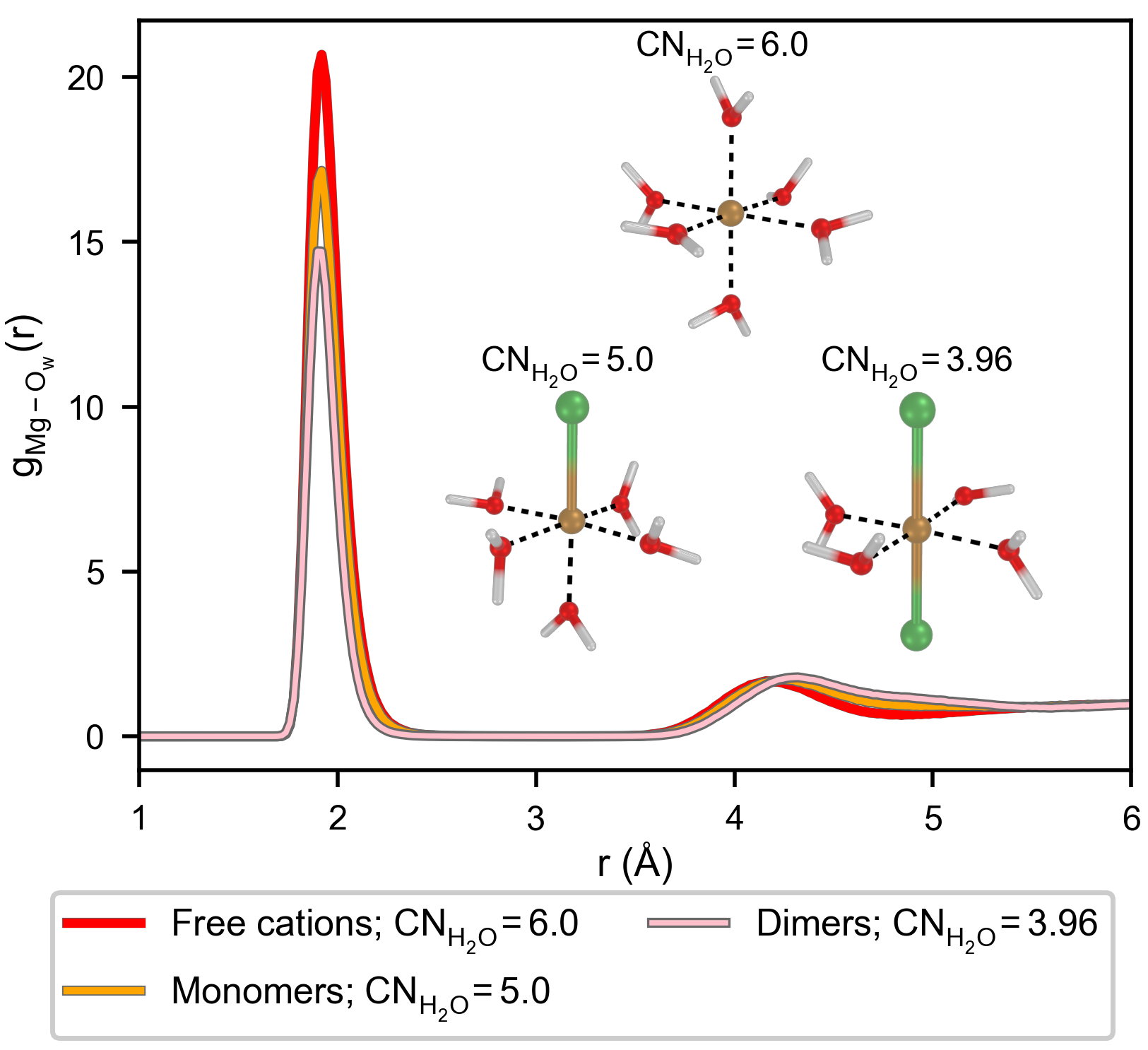}
    \caption{Radial distribution function between Mg$^{2+}$ and the oxygen atoms in water obtained from simulations using free ions (red curve), the monomer complex (yellow curve), the dimer complex (pink curve), and the water coordination
    number (CN$_{\ce{H_2O}}$) at $4$ m. 
    The total coordination number (including both the water molecules and chloride ions in the first solvation shell) is close to $6.0$ in all cases. 
    Insets depict typical solvation shells for free cations, monomers and dimers.}
    \label{fig:rdf}
\end{figure}

To assess the validity of the aforementioned simulation strategy, we analyzed the structures of the solvation shell around free cations, monomers and dimers.  
The radial distribution functions between \ce{Mg^{2+}} and $\mathrm{O_w}$, for free cations, monomers and dimers, at a concentration of 4 m are shown in Fig.~\ref{fig:rdf}. 
The expected solvation structure around both \ce{Mg^{2+}} and \ce{Fe^{2+}} is octahedral.~\cite{Ohtaki1993,Chaudhari2020}
If this structure is retained by the monomers and dimers, the total coordination number should be close to $6.0$.
Since we obtain hydration numbers of $5.0$ and $3.96$ for monomers and dimers, respectively, as shown in Fig.~\ref{fig:rdf}, we can conclude that constraining the cation-chloride distance does not 
%warp 
% significantly change
distort
the solvation shell around the cation.

\section{Theoretically expected viscosity trends of \ce{FeCl2} and \ce{MgCl2}}
\label{sec:lj_thought_experiment}

From Fig.2 in the main text, it is evident that while the viscosities of \ce{MgCl2} and \ce{FeCl2} are similar at low concentration, upto $2$ m, they diverge greatly at higher concentration.
From experiments, the viscosity of \ce{FeCl2} at $4$ m is $4.17$ mPa.s (Table 1 in the main text), compared to that of \ce{MgCl2} at the same concentration, equal to $4.73$ mPa.s (see also Sec.~\ref{sec:experimental_visc_MgCl2}). 
In the main text, we conclude that the formation of complexes (greater in \ce{FeCl2} compared to \ce{MgCl2}) is responsible for the observed viscosity trends.

Here, we rule out other possible explanations of the trends.
One could speculate that the difference in mass is responsible for the observed difference in viscosity, but there are several reasons to surmise that this is not actually the case.
The mass of the two most abundant components in both solutions, water and chloride, is the same.
The cation is the least abundant component, and the mass of \ce{Fe} is only $2.5$ times larger than that of \ce{Mg}. 

Furthermore, we approximate the \ce{FeCl2} and \ce{MgCl2} solutions as effective Lennard-Jones systems, for the sake of a qualitative argument that further illustrates our point.
Meier and coworkers\cite{viscosity_LJ} have shown that, in a Lennard-Jones (LJ) system, the KK and KU integrals are much smaller than the UU integral [see Eqs.~\eqref{shear_GK} and \eqref{pressure_tensor}], which is the main contribution. 
For such an LJ system, the viscosity, in reduced units, is universal (i.e., it does not depend on the masses) and it is given in units of $\eta^{*} = \eta \sigma^2 / \sqrt { m \epsilon }$.
In other words, for two systems with the same LJ parameters, the value of the viscosity is proportional to $\sqrt{m}$. 
The average mass of a particle (performing the average over molecules of water, cations and anions) in the  4 m \ce{FeCl2} solution is $22.32$ g/mol, whereas in the \ce{MgCl2} solution it is $20.46$ g/mol.
Thus, one would expect the viscosity of the \ce{FeCl_2} solution to be about $\sqrt{22.32/20.46}=1.05$ (i.e., $5\%$ larger) than that of \ce{MgCl_2}.
However, in the experiments, the viscosity of \ce{FeCl_2} is $12\%$ lower than that of \ce{MgCl_2}, resulting in a deviation from the expected result of about $17\%$. 
Consequently, we conclude that the mass is not responsible for the observed differences in experimental viscosity at high concentrations.
Moreover, we believe that the overestimation of viscosity obtained from our simulations for both \ce{MgCl_2} and \ce{FeCl_2} is not explained by a known tendency of the Madrid-2019 model to overestimate viscosity (see Sec.~\ref{sec:madrid_visc_overestimation} for more details).

\section{Predictions of speciation from chemical equilibrium models}
\label{sec:visual_minteq}

A popular (and free) program is Visual MINTEQ,~\cite{gustafsson2011visual} that enables the estimation of the population of complexes at 4 m for \ce{FeCl_2} and \ce{MgCl_2}.
This program uses the association constants (at infinite dilution) recommended by NIST, and obtains the equilibrium population after solving the equation of balance of mass, charge, iteratively, until all equilibrium constants and user-defined constraints (the concentration, pH, etc) are satisfied.
In principle, this seems like a reasonably sound strategy, but these calculations can be sensitive to 1) the association constants, and 2) the activity model used, such as the Specific Ion Theory\cite{sit_bronsted,sit_guggenheim,sit_ciavatta} (SIT), Davies equation\cite{Davies1938} , etc.
As detailed in the main text, experiments differ widely in the estimation of association constants.

The program clearly predicts complex formation in both \ce{MgCl_2} and \ce{FeCl_2} solutions. 
We emphasize that the quantitative complex populations predicted depend strongly on the chosen activity model.
However, independent of the activity model, Visual MINTEQ predicts higher association in \ce{MgCl2} compared to \ce{FeCl2}, which goes against the expectation of less complex formation in \ce{MgCl2}.\cite{DuboueDijon2017,Boehm2015, Luin2022}

\section{Densities from simulations}
\label{sec:densities}

\subsection{Simulations without complexes}

\begin{figure}[H]
  \includegraphics[width=0.6\linewidth]{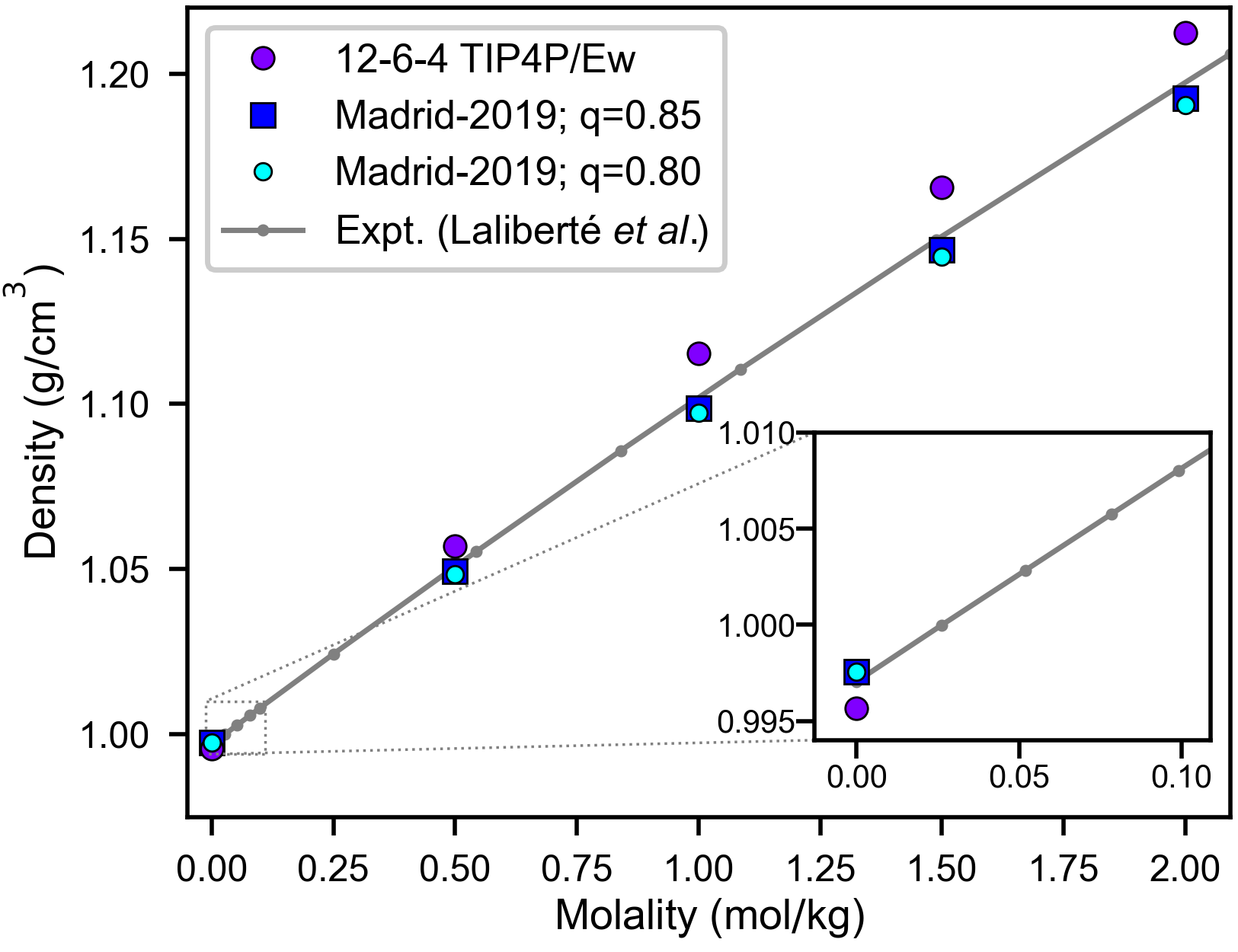}
  \caption{\label{fecl2_densities_comparison} Densities of \ce{FeCl_2} obtained from simulations, for low concentrations upto $2$ m, using the proposed Madrid-2019~\cite{zeron2019force} force-field for \ce{FeCl_2} (assuming free ions), and the 12-6-4 potential model of Ref.\cite{li2014taking}
  for \ce{Fe^{2+}} and TIP4P-Ew water along with with the Joung-Cheatham parameters~\cite{joung2008determination} for chloride-water interactions (again for TIP4P-Ew). Unlike the Madrid-2019 force field, cross-interactions are obtained using the Lorentz-Berthelot rules for the 12-6-4 potential. 
  Experimental densities in black are taken from~\citet{Laliberte2004} 
  The inset depicts a close-up view of the data for very low molalities, with different ranges on both axes. 
  The Madrid model with $q=0.80$ \textit{e} (cyan circles) uses the same parameters as those of the Madrid-2019 ($q=0.85$ \textit{e}) except for the values of $\sigma$ for the \ce{Cl-$\mathrm{O_w}$} and \ce{Mg-$\mathrm{O_w}$} interactions which are slightly different. 
  }
\end{figure}

In Fig.~\ref{fecl2_densities_comparison}, we present the results for the density of \ce{FeCl_2}, obtained from simulations using the proposed Madrid-2019 force-field, for low concentrations upto $2$ m, where complexes are not expected to form. 
We provide a comparison with a previously published model for \ce{Fe^{2+}} and water as described by the TIP4P/Ew\cite{Horn2004} model,  the 12-6-4 potential of~\citet{li2014taking}, using Joung-Cheatham parameters~\cite{joung2008determination} for the chloride -- water interactions (chloride parameters specifically designed for TIP4P-Ew) , and Lorentz-Berthelot mixing rules for cross-interactions. 
The densities obtained with the 12-6-4 potential show a systematic positive deviation of about $2\%$ from the experimental densities, for concentrations up to $2$ m.
The agreement of the TIP4P/Ew~\cite{Horn2004} and TIP4P/2005~\cite{Abascal2005} water models with experimental data for the density of pure water (i.e., at zero molality) is excellent, with deviations within $0.15\%$, as shown in the inset. 
It should be mentioned that the force-field for \ce{Fe^{2+}} from Ref.\cite{li2014taking} was designed for the cation at infinite dilution and was not tested at high concentrations of salt. 

\subsection{Simulations incorporating complexes}

\begin{table}[hbt!]
  \caption{Densities ($\rho$), in $\mathrm{g/cm^3}$ at $298.15$ K and $1$ bar, of \ce{FeCl_2} and \ce{MgCl2} at $4$ m, but with varying $\alpha$ (percentage of free cations in the system), calculated using the Madrid-2019 force field for both \ce{Fe^2+} and \ce{Mg^2+}. Note that $\alpha + \alpha' = 100$, since we assume that cations are either free or that they participate in monomers.} 
  \label{density_with_without_monomers}
  \small

  \begin{tabular}{lrrr}
      \mytoprule
      System & $\alpha$ & $\alpha'$ & $\rho$ \\
      \midrule 
      $\mathrm{MgCl_2}$ & 100 & 0 & 1.248 \\
      $\mathrm{MgCl_2}$ & 15 & 85 & 1.254 \\
      $\mathrm{MgCl_2}$ & 2.5 & 97.5 & 1.255 \\
      $\mathrm{FeCl_2}$ & 100 & 0 & 1.363 \\
      $\mathrm{FeCl_2}$ & 15 & 85 & 1.369 \\
      $\mathrm{FeCl_2}$ & 2.5 & 97.5 & 1.370 \\                                    
      \mybottomrule
  \end{tabular}
\end{table}

In this work, we also study the effect of complex speciation on the densities.
We define $\alpha$, $\alpha'$ and $\alpha''$ as the percentage of free cations, monomers and dimers, respectively.
Considering a mixture composed of only free cations and monomers, we compare the values of densities for a particular concentration ($4$ m) for both \ce{MgCl_2} and \ce{FeCl_2}, as shown in Table \ref{density_with_without_monomers}.
We conclude that the densities are relatively insensitive to complex formation, since the differences in densities correspond to a difference of about $0.5 \%$. 
We speculate that the density change caused by the inclusion of chloride ions in octahedral solvation shells (compared to a situation in which only water is in an octahedral solvation shell) is not significant. 

\section{Overestimation of viscosities by Madrid-2019}
\label{sec:madrid_visc_overestimation}

\begin{figure}[t]
    \includegraphics[width=0.6\linewidth]{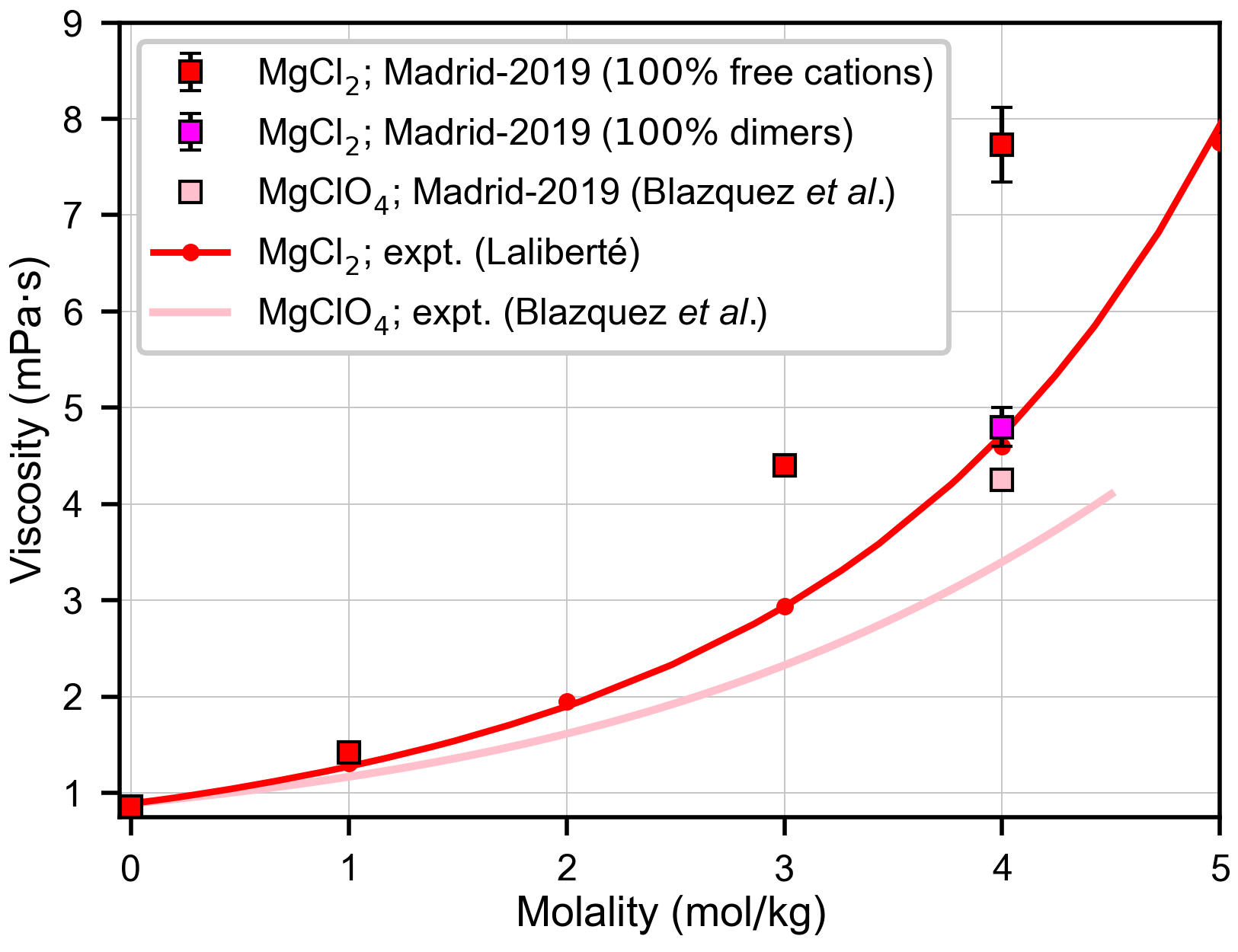}
    \caption{\label{fig:puzzle_si} Viscosities of \ce{MgCl_2} (solid red line) and \ce{Mg(ClO_4)_2} from experiments (solid pink line) and from simulations with the Madrid-2019 force-field (filled red and pink squares, respectively). Simulation 
    results at $4$ m are from this work. Experimental results for \ce{Mg(ClO_4)_2} are from~\citet{blazquez2025extending} Filled magenta square: simulation results for the Madrid-2019 force field of \ce{MgCl_2} at $4$ m assuming that all cations form \ce{[Mg(H_2O)_4Cl_2]}. 
    } 
\end{figure}

The Madrid-2019 model has been known to slightly overestimate viscosities.
However, we show that this overestimation cannot explain the large deviation from experimental viscosities for \ce{MgCl_2} and \ce{FeCl_2}.
For instance, in the case of \ce{NaCl}, the error in the prediction, at $4$ m, was $\approx0.4$ mPa$\cdot$s, whereas for \ce{MgCl_2} it is $\approx2.5$ mPa$\cdot$s. 
In the same vein, we observe that in the case of \ce{Mg(ClO_4)_2} (a solution which is somewhat similar to \ce{MgCl_2}) shown in Fig.~\ref{fig:puzzle_si}(b), the model overestimates the viscosity by about $1$ mPa$\cdot$s.
% -- but this deviation is much smaller than the deviation exhibited by \ce{MgCl_2} ($2.5$ mPa$\cdot$s).

We argue that complex formation in \ce{FeCl_2} and \ce{MgCl_2} solutions, at higher concentrations, causes the deviations in computed viscosities (from simulations wherein all cations are free). 
Artificially inserting complexes (such that all cations form "dimers" or \ce{[Mg(H2O)_4Cl_2]} complexes) brings the computed viscosity value closer to the experimental value, as shown in Fig.~\ref{fig:puzzle_si}. 

\section{Speciation effects on transport properties}
\label{sec:speciation_effects_transport}

\begin{table}[hbt!]
    \caption{Viscosity (in mPa$\cdot$s)  of \ce{FeCl_2} and \ce{MgCl_2} solutions at $4$ m obtained from simulations. 
    $\alpha$ denotes the percentage of free cations in the system (only monomers are considered so that $\alpha+\alpha'=100$). 
    The Yeh-Hummer correction~\cite{Yeh2004} is applied to the diffusion coefficients to account for finite-size effects. 
    } 
    \label{viscosity_with_without_monomers_main}
    \small
    \begin{tabular}{lccc}
        \mytoprule
        System & $\alpha$ & $D_{\ce{H_2O}}$ $\times 10^{9}$  ($\mathrm{m^2/s}$) & $\eta$ (mPa$\cdot$s)  \\
        \midrule
        \ce{MgCl_2} & $100$ & $0.366$ & $7.73(0.39)$ \\
        \ce{MgCl_2} & $15$  & $0.499$ & $5.71(0.33)$ \\
        \ce{MgCl_2} & $2.5$ & $0.522$ & $5.56(0.42)$ \\
        \ce{FeCl_2} & $100$ & $0.369$ & $7.68(0.41)$ \\
        \ce{FeCl_2} & $15$  & $0.496$ & $5.87(0.30)$ \\
        \ce{FeCl_2} & $2.5$ & $0.522$ & $5.54(0.32)$ \\
        \mybottomrule                             
    \end{tabular}
\end{table}

The viscosity and diffusion coefficient obtained in simulations for $4$ m solutions of \ce{FeCl_2} and \ce{MgCl_2}
are presented in Table \ref{viscosity_with_without_monomers_main}.

\begin{table}[hbt!]
  \caption{
  The product of the diffusion coefficient of water, and the viscosity of the solution ($D_{\ce{H_2O}} \times \eta$), reported in units of $10^{12} \times$ J/m, for \ce{FeCl_2} and \ce{MgCl_2} at $4$ m. 
  $\alpha$ denotes the percentage of free cations in the system.
  The Yeh-Hummer correction~\cite{Yeh2004} was applied to the diffusion coefficients (shown in units of $\mathrm{m^2/s}$ after being multiplied by $10^{9}$) to account for finite-size effects.
  Note that $\alpha + \alpha' = 100$, since we assume that cations are either free or that they participate in monomers.
  } 
  \label{tab:viscosity_with_without_monomers}
  \small
  \begin{tabular}{lcc}
      \mytoprule
      System & $\alpha$ & $D_{\ce{H_2O}} \times \eta$ \\
      \midrule
      \ce{MgCl_2} & 100 & 2.8 \\
      \ce{MgCl_2} & 15  & 2.8 \\
      \ce{MgCl_2} & 2.5 & 2.9 \\
      \ce{FeCl_2} & 100 & 2.8 \\
      \ce{FeCl_2} & 15  & 2.9 \\
      \ce{FeCl_2} & 2.5 & 2.9 \\
      \mybottomrule                             
  \end{tabular}
\end{table}

In Table \ref{tab:viscosity_with_without_monomers}, we present the product of the $D_{\ce{H_2O}}$ and the viscosity of the system.
This product is approximately constant and it is around 2.8(2) $10^{12}$ J/m.)
This is interesting, since it shows that a rough estimate of the viscosity can be obtained from the diffusion coefficient of water (a relatively cheap calculation). 

\begin{table}[hbt!]
  \caption{Diffusion coefficient ($\times 10^9 \mathrm{m^2/s}$) of the cations of \ce{FeCl_2} and \ce{MgCl_2} at 4 m  using the same force field for \ce{Fe^2+} and \ce{Mg^2+}. Results are shown for $\alpha=100$ (free ions, no monomers)  and for $\alpha=2.5$ (the majority of the cations is in the monomer form). 
  Note that $\alpha + \alpha' = 100$, since we assume that cations are either free or that they participate in monomers.
  } 
  \label{diffussion_of_cations}
  \small
  \begin{tabular}{l c c c c c}
  \mytoprule
  &  $\alpha=100$  &    $\alpha=2.5$   \\
  \mytoprule
  & $D_{\ce{M^{2+}}}$ & $D_{\ce{M^{2+}}}$, $D_{\ce{MCl^{+}}}$  \\
  \midrule
  \ce{MgCl_2} & 0.0955 & 0.15(5), 0.129(4) \\ 
  \ce{FeCl_2} & 0.0922 & 0.15(3), 0.130(2) \\
  \mybottomrule
\end{tabular}
\end{table}

The diffusion of the cations is shown in Table \ref{diffussion_of_cations}.
For $\alpha=100$, the diffusion of Fe is smaller than that of Mg (as expected from its higher mass). When monomers are formed, the diffusion coefficient of the cation increases significantly both for the free cation and for the cation forming the complex.

\section{Back-of-the-envelope calculation for viscosities}

\begin{table}[]
  \caption{ Changes in the viscosity $\eta$ in mPa$\cdot$s,  of $\ce{MgCl_2}$ 
  % (or \ce{FeCl2}: considering that both systems have the same viscosity values within error bars) 
  at the concentration of 4 m with the degree of association. 
  Calculations are performed with the Madrid-2019 force field using the $q=0.8 \ e$ model. 
  Results are presented for free ions $\alpha=100$, monomers $\alpha'=100$ and dimers 
  $\alpha''=100$} 
  \label{visc_comparison_monomer_dimer}
  \small
  \begin{tabular}{ccccc}
  \mytoprule
  System & $\alpha$ & $\alpha'$  &  $\alpha''$  & $\eta$    \\
  \midrule
  Free  ion         &  100  & 0    & 0     & $5.33(0.3)$   \\
  monomer complexes & 0    & 100   & 0     & $4.12(0.3)$  \\

  dimer complexes & 0     & 0   & 100        & $3.90(0.3)$  \\
  \mybottomrule
  \end{tabular}
\end{table}

In the main text, we proposed a simple relation between the viscosity with $100\%$ free ions, and the distribution of complexes ($\alpha$, $\alpha'$ and $\alpha''$)

Here, we estimate the coefficients of $\alpha'$ and $\alpha''$ in Eq (2) in the main text, denoted by $\delta'$ and $\delta''$, respectively (the rate of change of viscosity with $\alpha''$ or $\alpha'$).
Therefore, for the $q=0.85 \ e$ and $q=0.80 \ e$ Madrid-2019 models, we obtain $\delta'=0.020$ $\mathrm{mPa\cdot s}$ and $\delta'=0.010$ $\mathrm{mPa\cdot s}$, respectively, from the slopes of lines fitted to the data in Fig. 3 of the main text.

The value of $\delta''$ for the $q=0.85 \ e$ model can be obtained from:
\begin{equation}
    \begin{aligned}
        &\frac{ \eta_{(\alpha=100, \alpha'=0, \alpha''=0)} -  \eta_{(\alpha=0, \alpha'=0, \alpha''=100)} } {100} \\
        = &\frac{7.73 - 4.80 } {100} \ \mathrm{mPa\cdot s} \\ 
        \approx &0.029 \ \mathrm{mPa\cdot s},
   \end{aligned}
\end{equation}
where the viscosity values were obtained for the $q=0.85$ Madrid-2019 model.

The value of $\delta''$ for the $q=0.8 \ e$ model can be obtained from:
\begin{equation}
    \begin{aligned}
        &\frac{ \eta_{(\alpha=100, \alpha'=0, \alpha''=0)} -  \eta_{(\alpha=0, \alpha'=0, \alpha''=100)} } {100} \\
        = &\frac{5.33 - 3.90} {100} \ \mathrm{mPa\cdot s} \\ 
        \approx &0.014 \ \mathrm{mPa\cdot s},
   \end{aligned}
\end{equation}
where the viscosity values were obtained for the $q=0.8 \ e$ Madrid-2019 model, collected in Table~\ref{visc_comparison_monomer_dimer}.

Furthermore, we provide justifications for the values taken for $\log K_1$ and $\log K_2$.
Values of $\log K_1$) (i.e., logarithm to base 10) and $\log K_2$ for some divalent cations with many ligands can be found in the monumental and trusted Lange's Handsbook.\cite{lange}
For \ce{Fe^{2+}}, the handbook only reports $\log K_1$ and refers to the value of \citet{olerup} (equal to $0.36$).
However, the handbook does provide the values of $\log K_1$ and $\log K_2$ for \ce{Fe^{3+}}, equal to $1.48$ and $0.65$, respectively, as well as those for \ce{Pb^{2+}}, equal to $1.62$ and $0.82$, respectively.
Of course, the individual values depend on the cation, but we observe that the approximation $\log K_2 \approx \log K_1 -0.8$  seems reasonable.
Intuitively, this makes sense since the \ce{Cl^{-}}-\ce{Cl^{-}} electrostatic repulsion would make the formation of the second complex somewhat less favorable (that means that the second equilibrium constant is similar to the first and differs from it, at most, by an order of magnitude).
In this experiment, we shall set all activity coefficients to 1 (we stress, once again, that we are just attempting to do a qualitative calculation).
By assuming the value $-0.20$ for $\log K_1$ from NIST, and $-1.0$ for $\log K_2$, we obtain (for a $4$ m solution of \ce{FeCl_2}): 
$\alpha=22$, $\alpha'=56$ and $\alpha''= 22$ (Visual MINTEQ with the SIT model, and without dimer formation included, would lead to $\alpha=29$, which is not so different from our value).
By replacing the value $\eta^0$ for the model with $q=0.8$ (i.e $5.33$ mPa$\cdot$s for $\alpha=100$ ), and using Eq.~(6) from the main text with these values of $\alpha$, $\alpha'$ and $\alpha''$, one obtains $4.46$ mPa$\cdot$s for the viscosity of \ce{FeCl_2}, which is not so different from the experimental value measured in this work (equal to $4.17$ mPa$\cdot$s; see Table 1 in the main text).
By assuming a $\log K_1$ of $-1.1$ for \ce{MgCl_2} (i.e., $0.9$ below the value of \ce{FeCl_2}), and $\mathrm{log}K_2=-1.1-0.8=-1.9$ we obtain, for a $4$ m solution of \ce{MgCl_2}:
$\alpha=64$, $\alpha'=33$, $\alpha''=3$.
Similarly, we obtain $4.96$ mPa$\cdot$s for the viscosity of \ce{MgCl_2}, which is very close to the viscosity of \ce{MgCl_2} from experiments (as shown in Fig.~\ref{fig:mgcl2_expt_chaos}).

%%%%%%%%%%%%%%%%%%%%%%%%%%%%%%%%%%%%%%%%%%%%%%%%%%%%%%%%%%%%%%%%%%%%%
%% The appropriate \bibliography command should be placed here.
%% Notice that the class file automatically sets \bibliographystyle
%% and also names the section correctly.
%%%%%%%%%%%%%%%%%%%%%%%%%%%%%%%%%%%%%%%%%%%%%%%%%%%%%%%%%%%%%%%%%%%%%

\section{ Evaluation of the Jones-Dole coefficient of a force-field }

The Jones-Dole equation describes the variation of the viscosity of electrolyte solutions quite well, according to the relation:
\begin{equation}
  \eta / \eta_w = 1 + A \sqrt m +  B m + Cm^2 + ...
  \label{eq:jones_dole}
\end{equation}
where $\eta$ is the viscosity of the solution at molality m and $\eta_w$ is the viscosity of pure water. The coefficient $A$ is always positive and rather small and is at least one order of magnitude smaller than $B$. 
The term $A$ captures the viscosity change due to Coulombic ion-ion interactions.
The term $B$ originates from from ion-water interactions, and the term $C$ represents the interactions between hydrated ions and the formation of complexes.
Determining $A$ and $B$ from simulations is a \textit{tour of force}, as it requires extremely long simulations to estimate the viscosity with very high accuracy at low concentrations. 
In fact, there is only one paper in the literature \cite{yue2019dynamic} which rigorously evaluates the Jones-Dole coefficients for certain popular force-fields of 1:1 electrolytes. 
In that work, $B$ coefficients were obtained by fitting the simulation results for concentrations between $0.1$ and $1$ m, while the values of $A$ were not reported, since the associated statistical error was larger than the absolute values. 
In this work, we estimate the $B$ coefficients by using the simple expression (obtained from Eq.~\eqref{eq:jones_dole} after neglecting the quadratic term):
\begin{equation}
B' \simeq ( \eta / \eta_w  - 1  - A \sqrt{m} )/ m
  \label{B_approximate}
\end{equation}
Given that $A$
$<$$<$ $B$, we use the experimental values of $A$ to obtain $B'$. 
The value of $A$ is is  $\approx$ $0.02$ (kg/mol)$^{1/2}$ for 2:1 electrolytes \cite{marcusb}.
Furthermore, the results of \citet{yue2019dynamic} strongly suggest that the values of $A$ of atomistic force-fields are very close to the experimental ones. 
To evaluate $B'$, a concentration which is low enough that the quadratic term contribution in the Jones-Dole equation (Eq.~\eqref{eq:jones_dole}) can be neglected is needed. 
We shall use $0.6$ m for \ce{MgCl_2} and \ce{FeCl_2}. 
In fact, if one uses experimental results\cite{laliberte2007model} for $\eta$ and $\eta_w$ at $0.6$ m for \ce{CaCl_2} and \ce{MgCl_2} in Eq.~\eqref{B_approximate}, one obtains a value of $B'$ that is only $0.01$ kg/mol higher than the experimental value of $B$ reported by Marcus\cite{marcusb,book-marcus} for these salts ($0.01$ kg/mol is also the experimental uncertainty of $B$). 
The notation $B'$ just reflects that $B'$ (as given by Eq.~\eqref{B_approximate}) is an approximation of the true value of $B$ (although it is a good approximation as when implemented with experimental values of viscosities since it deviates by only $0.01$ kg/mol from the exact value of $B$).

In order determine the coefficient $B'$, we performed $24$ independent simulations (at $0.6$ m) in the \textit{NVT} ensemble (keeping the rest of the simulation details the same as those specified in Sec.~\ref{subsec:sim_details}) of 20 ns each (i.e., we used a total simulation time of $480$ ns to determine the viscosity of one salt and force-field). 
The volume used in the \textit{NVT} simulations was obtained from a \textit{NpT} run of $20$ ns. 
The system contains $4440$ molecules of water and the number of ions needed for the required concentration, i.e., $0.6$ m. 
Our estimated uncertainty for $B'$ is $0.03$ kg/mol (obtained from the uncertainties of the viscosities of the solutions and that of pure water plus that introduced by Eq.~\eqref{B_approximate}). 
The viscosity of TIP4P/2005 water at $298.15$ K and $1$ bar is $0.866(3)$ mPa·s (the experimental one is $0.890$ mPa·s). 
Using the force-field described in the main text for the scaled charge $0.8$, we obtained (at $0.6$ m and $298.15$ K and 1 bar) a viscosity of $1.120(4)$ mPa.s for \ce{FeCl_2} and of $1.097(4)$ for \ce{MgCl_2}. Therefore, we obtained $B'=0.46(3)$ kg/mol for \ce{FeCl_2} and $B'=0.42(3)$ kg/mol for \ce{MgCl_2}.

%%%%%%%%%%%%%%%%%%%%%%%%%%%%%%%%%%%%%%%%%%%%%%%%%%%%%%%%%%%%%%%%%%%%%
%% The appropriate \bibliography command should be placed here.
%% Notice that the class file automatically sets \bibliographystyle
%% and also names the section correctly.
%%%%%%%%%%%%%%%%%%%%%%%%%%%%%%%%%%%%%%%%%%%%%%%%%%%%%%%%%%%%%%%%%%%%%
\putbib

\end{bibunit}

\end{document}